\documentclass[12pt,preprint,3p,times]{elsarticle}
\usepackage{lineno,hyperref}
\usepackage{amsmath}
\usepackage{kotex}
\usepackage{multirow}
\usepackage{graphicx}
\usepackage{amssymb}
\usepackage{relsize}
\usepackage{color}
\usepackage[LGR,T1]{fontenc}
\usepackage{bigstrut}
\usepackage{tikz}
\usepackage{verbatim}
\usepackage{float}
\usepackage{subfigure}
\usepackage{adjustbox}
\usepackage{caption}
\usepackage{setspace}
\usepackage{makecell}
\usepackage[figuresright]{rotating}
\usepackage{lscape}
\usepackage[toc,page]{appendix}
\usepackage{siunitx}
\journal{Arxiv}
\begin{document}
%
%
\begin{frontmatter}

\title{Implicit inverse force identification method of acoustic liquid-structure interaction finite element model}

\author[a]{ Seungin Oh }
\author[a]{ Chang-uk Ahn}
\author[b]{ Kwanghyun Ahn }
\author[a]{ Jin-Gyun Kim \corref{mycorrespondingauthor}}

\cortext[mycorrespondingauthor]{Corresponding authors}
\ead{ jingyun.kim@khu.ac.kr }
\address[a]{ Department of Mechanical Engineering (Integrated Engineering), Kyung Hee University, 1732, Deogyeong-daero, Giheung-gu, Yongin-si, Gyeonggi-do 17104, Korea }
\address[b]{ Korea Atomic Energy Research Institute, 111, Daedeok-daero 989beon-gil, Yuseong-gu, Daejeon, Korea }

%
%
\begin{abstract}
The two-field vibroacoustic finite-element (FE) model requires a relatively large number of degrees of freedom compared to the monophysics model, and the conventional force identification method for structural vibration can be adjusted for multiphysics problems. In this study, an effective inverse force identification method for an FE vibroacoustic interaction model of an interior fluid-structure system was proposed. The method consists of: (1) implicit inverse force identification based on the Newmark-$\beta$ time integration algorithm for stability and efficiency, (2) second-order ordinary differential formulation by avoiding the state-space form causing large degrees of freedom, (3) projection-based multiphysics reduced-order modeling for further reduction of degrees of freedom, and (4) Tikhonov regularization to alleviate the measurement noise. The proposed method can accurately identify the unmeasured applied forces on the in situ application and concurrently reconstruct the response fields. The accuracy, stability, and computational efficiency of the proposed method were evaluated using numerical models and an experimental testbed. A comparative study with the augmented Kalman filter method was performed to evaluate its relative performance.
\end{abstract}
\end{frontmatter}


%
%
\section{Introduction}\label{section1}

The structural dynamics of liquid fluids subjected to vibration and/or transient excitations (e.g., launcher payloads, nuclear power plants, tanker ships, and automobiles) are important for aerospace engineering, nuclear engineering, civil engineering, mechanical engineering, and naval architecture.
Various analytical (and semi-analytical) methods have been proposed to describe the dominant features of fluid-filled tankers by combining the structural mode shapes and acoustic fluid potential~\cite{Amabili1996, Leissa1973, Kwak2000, Amabili2003}. 
Finite-element (FE) modeling is one of the most popular numerical methods  to design and analyze the interior fluid-structure interaction  problem of modern structures with complex geometries \cite{Ohayonbook}. 
The two-field coupled FE model is defined as a discretized second-order ordinary differential equation, and the system matrices (mass, stiffness, and damping) defined numerically are adjusted by conventional FE model updating techniques with experimentally measured data such as natural frequencies, damping ratios, and a frequency response function ~\cite{Friswell,Rates_Fox,ma1991sensitivity,dhandole2010comparative}. 

Force identification is also a primary process for achieving a precise FE model. However, in practice, direct measurement of the applied forces is difficult or impossible owing to various limitations.
Therefore, inverse force identification methods have been studied to overcome this issue in terms of structural dynamics and vibration.
Law et al.~\cite{lau1997} proposed the time domain force identification method using explicit integration method to identify the moving forces.
Zhu and Law implemented a method for multi-span bridges to identify moving forces\cite{Zhu2000,Zhu2001}.
In addition, Law and Fang proposed a state-space-representation-based identification method~\cite{Law2001} using a dynamic programming technique.
Later, Kammer proposed an inverse force identification method based on a set of inverse Markov parameters~\cite{kammer1998input}, and Law et al. also used a Markov parameter system to identify time-varying wind loads.~\cite{Law2005}.
 The Kalman filter technique is also implemented in inverse force identification methods to restrain the effect of inherent noise in the measured signal.
Lourens et al. proposed a structural force identification method using the augmented Kalman filter (AKF) algorithm\cite{lourens2012augmented}, and Naets et al. suggested a stable force identification process by adding dummy measurements at the position level\cite{naets2015stable}. Azam et al. implemented the dual Kalman filter concept to prevent numerical issues attributed to the unobservability and rank deficiency of input signals\cite{azam2015dual}. Recently, Kalman-filter-based methods have been extended to nonlinear systems by implementing an unscented Kalman filtering algorithm on the structural systems\cite{erazo2017offline}. These Kalman-filter-based methods are also based on the state-space representation of the numerical model and explicit time-integration methods.

These methods can successfully identify the unknown applied forces of the structural dynamics and vibration systems in the time domain. The problem is that inverse force identification has rarely been studied for the FE-based vibroacoustic interaction model of structural vibration containing interior liquid fluid, and conventional force identification methods can be adjusted based on the following aspects.
(1) In structural vibration, the fluid domain idealized as a linear acoustic fluid is discretized using fluid pressure, velocity potentials, displacement potentials, etc., which implies that the number of fluid degrees of freedom (DOFs) is added to the empty tanker structure model. Therefore, the computational cost issue with conventional inverse force identification is severe in multiphysics problems.
(2) Most existing time-domain force identification methods are based on the state-space form. 
It would be resolved with a reduced order modeling in linear structural vibration~\cite{Craig1968, Kim2015, Kim2017}, but the state-space form with the virbo-acoustic multiphysics model reduction is another difficulty \cite{Ohayonbook,kim2019strongly,kim2020multiphysics,Peter2016}. 
This can be resolved by reduced-order modeling in linear structural vibration, but the state-space form with vibro-acoustic multiphysics model reduction is another difficulty. 
(3) Explicit algorithms and Kalman filtering within the state-space form have been successfully and widely used for inverse force identification~\cite{lourens2012augmented, azam2015dual, naets2015stable}. 
This is sufficient for rigid-body system dynamics and a simplified structural dynamics model with relatively low dominant natural frequencies, such as high-rise towers, multistorey buildings, bridges, frames, and truss structures. However, the disadvantages of a large discretization error with a low sampling frequency or a long sampling duration are well known; hence, Liu et al. proposed an explicit form of an implicit formulation for inverse force identification in structural dynamics\cite{liu2014explicit}. 
The Tikhonov regularization method was then employed to cancel the noise within the Newmark-$\beta$ algorithm instead of the Kalman filter. 
(4) Several researchers have addressed inverse force identification for vibroacoustic systems~\cite{vaitkus2019application, paxton1996experimental, lopp2019bayesian}.
. However, these were based on frequency-domain-based approaches and were only studied for noise source identification for NVH problems, such as cabin noise and room noise. This study was motivated by these aspects.

The aim of this study is to develop an efficient inverse force identification method for the structural dynamics of a liquid fluid subjected to vibration and/or transient excitations. The proposed inverse force identification method is based on the time-domain approach, which is capable of indirectly identifying the excitation forces using a limited amount of sensing information such as displacement and acceleration. 
It is then derived from the Newmark-$\beta$ time-integration algorithm in an implicit manner to achieve an unconditionally stable condition and computational efficiency with a relatively large time step. 
The second-order ordinary differential formulation is directly used instead of the state-space form to avoid an increasing number of DOFs of the vibroacoustic formulation. In this work, the well known displacement $(\textit{u})$-pressure $(\textit{p})$  formulation is considered to describe the interaction between the elastic structure and acoustic liquid fluid.
The Tikhonov regularization technique\cite{tikhonov1963solution,tikhonov1995numerical} is then implemented to alleviate the measurement noise and control the ill-posedness of the numerical process within the Newmark-$\beta$ method.
We also pursue an accurate and efficient reconstruction of the unmeasured response data (displacement, stress, strain, pressure, etc.) in the entire domain as well as the applied force identification.
Projection-based multiphysics reduced-order modeling is then employed in the proposed force identification method~\cite{kim2019strongly,kim2020multiphysics}. 
The performance of the proposed method is evaluated using numerical examples of an h-shaped pipe structure containing a liquid fluid. A numerical test is designed to investigate simple sinusoidal and random loads with different noise levels. Some measured response information is used to identify the unmeasured forces.
Comparison studies of both accuracy and efficiency are performed using the AKF method~\cite{lourens2012augmented}.
For the experimental study, a test bed is set up with an L-shaped pipeline filled with liquid water. For the experimental test, a single-point displacement signal is used as the source displacement for the proposed inverse force identification process. The accuracy, stability, and computational efficiency of the proposed method are then evaluated by comparing the identified results with reference quantities.

The remainder of this paper is organized as follows: In Section.~\ref{section2}, a general description of the inverse dynamic problem, the vibroacoustic reduced-order modeling, and the proposed implicit inverse force identification process is presented.
The detailed numerical test conditions and results are discussed in Section.~\ref{section3}, 
The experimental test conditions and results are described in Section.~\ref{section4}. 
The conclusions are presented in Section.~\ref{section5}.

%
%
\section{Problem formulation}\label{section2}
\subsection{Problem statements}
In this study, an identification process for the unmeasured applied forces on a vibroacoustic system is proposed. The force identification process can be defined as a minimization problem for determining the proper force values that can cause the measured physical responses of the target systems. Furthermore, the implied measurement errors in the measured physical quantities can significantly pollute the identified results owing to the ill-posedness of the inverse problem. Tikhonov regularization ~\cite{tikhonov1963solution,tikhonov1995numerical} is a widely used regularization method to solve such types of inverse dynamics problems. In this method, considering the measurement errors of the measured quantities, the following damped least-squares error equation can be defined:

    \begin{equation}\label{eq:error}
        \mathbf{J}(\mathbf{f},\alpha)=\left\| {\mathbf{z}}_{m}-\mathbf{z}_{n}\right\|^{2}+\alpha \left\|  {\mathbf{f}} \right\| ^{2},
    \end{equation}
    where $\mathbf{J}$ is the damped least-squares error and $\alpha$ is the regularization parameter. The stability and accuracy of the regularized solution can be controlled by changing the parameter $\alpha$, which can be obtained using various parameter estimation algorithms such as the L-curve method\cite{hansen1998rank}.
    $\mathbf{z}$ is the input response vector, and subscripts $m$ and $n$ denote the measured and numerically predicted values, respectively. $\mathbf{f}$ is an unknown applied force vector. In practice, the system response can only be measured at a few points. Hence, the response information from the measurement point can be selected from the numerical prediction values using the following relationship:
    
\begin{equation}\label{eq:Selection}
    {}^{t+\Delta t}\mathbf{z}_{n}=\left[ \begin{array}{c c c} \mathbf{S}_{d} & \mathbf{0} & \mathbf{0}\\ \mathbf{0} & \mathbf{S}_{v} & \mathbf{0}\\\mathbf{0} & \mathbf{0} & \mathbf{S}_{a} \end{array} \right] \left[ \begin{array}{c} \mathbf{d} \\ \dot{\mathbf{d}} \\ \ddot{\mathbf{d}} \end{array} \right],
\end{equation}
where $\mathbf{S}$ is the selection matrix for choosing the measured response vector of the numerical model, and the subscripts $d$, $v$, and $a$ denote the selection matrices for response vector and its first and second derivatives $(\mathbf{S}_{d} \in {\mathbb{R}}^{{n_{zd}}\times {n_{d}}},\mathbf{S}_{v} \in {\mathbb{R}}^{{n_{zv}}\times {n_{v}}}\mathbf{S}_{a} \in {\mathbb{R}}^{{n_{za}}\times {n_{a}}})$. The selection matrices can be constructed in the form of a boolean matrix that specifies the locations of the applied force elements. 
    To define the response vectors ($\mathbf{d}, \dot{\mathbf{d}}, \ddot{\mathbf{d}}$) of the system, an appropriate numerical model should be constructed. In this study, the FE model is used to describe the multiphysical properties of vibroacoustic systems. The $(\mathbf{u},\mathbf{p})$ formulation is a classical method of generating an FE model of vibroacoustic systems. In the formulation, mechanical properties of both displacement ($\mathbf{u}$) and pressure ($\mathbf{p}$) fields of the vibroacoustic system can be described while considering their coupling effects. The undamped $(\mathbf{u},\mathbf{p})$ formulation of the vibroacoustic system could be defined by the following equations of the motion: 
\begin{subequations}\label{eq:eom}
\begin{align}
    &\mathbf{A}\ddot{\mathbf{d}}+\mathbf{B}{\mathbf{d}}=\mathbf{S}_{f}\mathbf{f},\\
    &\mathbf{A}=\left[\begin{array}{cc}
        \mathbf{M}^{s} & \mathbf{0} \\
        \rho^{f} c^{2} \mathbf{C}^{T} & \mathbf{M}^{f} 
    \end{array} \right],~
    \mathbf{B}=\left[\begin{array}{cc}
        \mathbf{K}^{s} & -\mathbf{C} \\
        \mathbf{0} & \mathbf{K}^{f} 
    \end{array} \right],~
    \mathbf{d}=\left[ \begin{array}{c}
        \mathbf{u}  \\
        \mathbf{p}
    \end{array}  \right],~
    \mathbf{S}_{f}\mathbf{f}=\left[ \begin{array}{c}
        \mathbf{f}^{s}  \\
        \mathbf{f}^{f}
    \end{array}  \right],
    \end{align}
\end{subequations}
where matrices $\mathbf{A}$, $\mathbf{B}$, and $\mathbf{C}$ are the mass, stiffness, and coupling matrices of the vibroacoustic system, respectively. $\rho^{f}$ and $c$ are the density and speed of sound values of the system, respectively. $\mathbf{S}_{f}$ is the selection matrix of the force vector that connects the applied force elements to the equations of motion of the target system $(\mathbf{S}_{f} \in {\mathbb{R}}^{{n_{d}}\times {n_{f}}})$.
The superscripts $s$ and $f$ denote the structure and fluid components of the system, respectively.

\subsection{Reduced-order modeling}

The proposed identification process includes a sequential time integration process using a generated vibroacoustic FE model. The massive DOF count of the practical FE model could disturb the efficient computation of the proposed process. Hence, a proper reduced-order modeling technique should be implemented to reduce the computational burden during the identification process. 
Considering the arbitrary projection-based reduced-order modeling method and assuming $\hat{\mathbf{T}}$ is the transformation matrix of the method, the entire system can be reduced by multiplying the transformation matrix by Eq.~\eqref{eq:eom} as

\begin{subequations}\label{eq:eom_reduced}
    \begin{align}
    \hat{\mathbf{A}} \ddot{\hat{\mathbf{d}}}+
    \hat{\mathbf{B}} {\hat{\mathbf{d}}}=\hat{\mathbf{f}},\\
    \hat{\mathbf{A}}=\hat{\mathbf{T}}^{T}\mathbf{A}\hat{\mathbf{T}},~
    \hat{\mathbf{B}}=\hat{\mathbf{T}}^{T}\mathbf{B}\hat{\mathbf{T}},~
    \hat{\mathbf{f}}=\hat{\mathbf{T}}^{T}\mathbf{S}_{f}\mathbf{f},~\mathbf{d}\approx\hat{\mathbf{T}}\hat{\mathbf{d}}.
    \end{align}
\end{subequations}

 Equation~\eqref{eq:Selection} can then be approximated using the reduced response vectors ($\hat{\mathbf{d}}$, $\dot{\hat{\mathbf{d}}}$, $\ddot{\hat{\mathbf{d}}}$) and their transformation matrix ($\hat{\mathbf{T}}$) as

\begin{equation}\label{eq:Selection_red}
    {}^{t+\Delta t}\mathbf{z}_{n} \approx \hat{\mathbf{S}}
    \left[ \begin{array}{c} \hat{\mathbf{d}} \\ \dot{\hat{\mathbf{d}}} \\ \ddot{\hat{\mathbf{d}}} \end{array} \right],~\hat{\mathbf{S}}=\left[ \begin{array}{c c c} \mathbf{S}_{d}\hat{\mathbf{T}} & \mathbf{0} & \mathbf{0}\\ \mathbf{0} & \mathbf{S}_{v}\hat{\mathbf{T}} & \mathbf{0}\\\mathbf{0} & \mathbf{0} & \mathbf{S}_{a}\hat{\mathbf{T}} \end{array} \right].
\end{equation}

In this study, a strongly coupled vibroacoustic model reduction method\cite{kim2019strongly} is implemented to reduce the computational cost. The method considers the strong connection between the structure and fluid; therefore, it can dramatically reduce the DOFs while preserving the vibration properties of the original system. The strongly coupled vibroacoustic reduction technique is implemented to alleviate the computational costs in this research; however, any other projection-based vibroacoustic model order reduction technique could be used to enhance the efficiency and/or accuracy of the proposed force identification process. The detailed derivation process of the strongly coupled reduction technique can be found in the referred paper \cite{kim2019strongly}, and a simplified process is listed in \ref{appendix:A}.
Using this method, the transformation matrix can be defined as

\begin{equation}\label{eq:Trans_total}
    \hat{\mathbf{T}}=\left [ \begin{array}{cc}
        \boldsymbol{\Phi}_{d} & \boldsymbol{\Psi}\tilde{\boldsymbol{\Xi}}_{d} \\
        \mathbf{0} & \tilde{\boldsymbol{\Xi}}_{d}
    \end{array} \right].
\end{equation}

Meanwhile, to consider the dynamic response of the system, the damping matrix $\hat{\mathbf{D}}$ of the system can be added using the classical Rayleigh damping theory as
\begin{subequations}\label{eq:eom_reduced_damped}
    \begin{align}
    \hat{\mathbf{A}} \ddot{\hat{\mathbf{d}}}+\hat{\mathbf{D}}\dot{\hat{\mathbf{d}}}+
    \hat{\mathbf{B}} {\hat{\mathbf{d}}}=\hat{\mathbf{f}},\\
    \hat{\mathbf{D}}=\left[  \begin{array}{cc}
        \hat{\mathbf{D}}^{s} & \mathbf{0} \\
        \mathbf{0} & \hat{\mathbf{D}}^{f}
    \end{array} \right],~
    \hat{\mathbf{D}}^{s}=a^{s}_{1}\mathbf{I}_{d}^{s}+a^{s}_{2}\boldsymbol{\Lambda}_{d}^{s},~
    \hat{\mathbf{D}}^{f}=a^{f}_{1}\mathbf{I}_{d}^{f}+a^{f}_{2}\tilde{\boldsymbol{\Gamma}}_{d}^{f},
    \end{align}
\end{subequations}
where matrices $\boldsymbol{\Lambda}$ and $\tilde{\boldsymbol{\Gamma}}$ are the chosen dominant eigenvalue matrix of the structure and fluid part of the reduced system, respectively. Coefficients $a_{1}$ and $a_{2}$ are the mass and stiffness proportional damping coefficients of the system, which are independently defined for each structure and fluid part of the system, respectively. The defined equations of motion of the target system are then used as the numerical model, and the proposed identification process is derived based on the generated model.

\subsection{Implicit inverse force identification}
     To define the relationship between applied forces $\mathbf{f}$ and response vectors ($\hat{\mathbf{d}},\dot{\hat{\mathbf{d}}},\ddot{\hat{\mathbf{d}}}$), Eq.~\eqref{eq:eom_reduced} can be discretized using a numerical differentiation. In this study, the well-known Newmark-$\beta$ time integration method~\cite{newmark1959method} is used to discretize the equations of motion.
     From the Newmark-$\beta$ method, the implicit force equilibrium equation of the reduced system can be described as follows:
     \begin{equation} \label{eq:eom_implicit}
    \hat{\mathbf{A}} {}^{t + \Delta t}\ddot{\hat{\mathbf{d}}}+\hat{\mathbf{D}}{}^{t + \Delta t}\dot{\hat{\mathbf{d}}}+
    \hat{\mathbf{B}} {}^{t + \Delta t}{\hat{\mathbf{d}}}={}^{t+\Delta t}\hat{\mathbf{f}}.
\end{equation}

    From the Newmark-$\beta$ method, the velocity $\dot{\hat{\mathbf{d}}}$, and acceleration $\ddot{\hat{\mathbf{d}}}$ vectors can be derived as
    
\begin{subequations}\label{eq:response_t}
        \begin{align}
        {}^{t+\Delta t}\dot{\hat{\mathbf{d}}}&=\frac{\partial{}^{t+\Delta t}\hat{\mathbf{d}}}{\partial t}
        ={}^{t}\dot{\hat{\mathbf{d}}}+\Delta t(1-\delta){}^{t}\ddot{\hat{\mathbf{d}}}+\delta \Delta t{}^{t+\Delta t}\ddot{\hat{\mathbf{d}}},\\
        {}^{t+\Delta t}\ddot{\hat{\mathbf{d}}}&=\frac{\partial {}^{t+\Delta t}\dot{\hat{\mathbf{d}}}}{\partial t}
        =\frac{1}{\beta \Delta t^{2}}({}^{t+\Delta t}\hat{\mathbf{d}}-{}^{t}\hat{\mathbf{d}})-\frac{1}{\beta \Delta t}{}^{t}\dot{\hat{\mathbf{d}}}-(\frac{1}{2 \beta}-1){}^{t}\ddot{\hat{\mathbf{d}}},
            \end{align}
        \end{subequations}
        where $\delta$ and $\beta$ are the time integration coefficients of the Newmark-$\beta$ method and $\Delta t$ is the time increment value used.
        Substituting Eq.~\eqref{eq:response_t} into Eq.~\eqref{eq:eom_implicit} and rearranging, the discretized equations of motion can be redefined as
   
    \begin{subequations}\label{eq:implicit_explicitform_rom}
    \begin{align}
        \hat{\mathbf{K}}{}^{t+\Delta t}\hat{\mathbf{d}}&={}^{t+\Delta t}\hat{\mathbf{f}}+{}^{t+\Delta t}\hat{\mathbf{r}},\\
        \hat{\mathbf{K}}&=\hat{\mathbf{B}}+\hat{\mathbf{A}}\left(\frac{1}{\beta \Delta t^{2}}\right)+\hat{\mathbf{D}}\left( \frac{\delta}{\beta \Delta t} \right),\\
        {}^{t+\Delta t}\hat{\mathbf{r}} &= \hat{\mathbf{A}}\left [\frac{1}{\beta \Delta t^{2}}{}^{t}\hat{\mathbf{d}}+\frac{1}{\beta \Delta t}{}^{t}\dot{\hat{\mathbf{d}}}+(\frac{1}{2\beta}-1){}^{t}\ddot{\hat{\mathbf{d}}}\right ]
        +\hat{\mathbf{D}}\left [(\frac{\delta}{\beta \Delta t}){}^{t}\hat{\mathbf{d}}+(\frac{\delta}{\beta}-1){}^{t}\dot{\hat{\mathbf{d}}}+(\frac{\delta \Delta t}{2\beta}-\Delta t){}^{t}\ddot{\hat{\mathbf{d}}} \right ].
    \end{align}
    \end{subequations}
    
    Using Eq.~\eqref{eq:implicit_explicitform_rom}, the response vector can be computed as
    
    \begin{equation}\label{eq:response_numerical_total}
        {}^{t+\Delta t}\hat{\mathbf{d}}= \hat{\mathbf{H}}\hat{\mathbf{T}}^{T}\mathbf{S}_{f}{}^{t+\Delta t} {\mathbf{f}}+{\hat{\mathbf{H}}} {}^{t+\Delta t}\hat{\mathbf{r}},~
        \hat{\mathbf{H}}=\hat{\mathbf{K}}^{-1}.
    \end{equation}
    
    Both the first and second derivatives of the response vector can then be derived by substituting Eq.~\eqref{eq:response_numerical_total} into Eq.~\eqref{eq:response_t} as
    
    \begin{subequations}\label{eq:response_numerical_derivatives}
    \begin{align}
        {}^{t+\Delta t}\dot{\hat{\mathbf{d}}} &= \frac{\delta}{\beta \Delta t} \left( \left[ \hat{\mathbf{H}} \hat{\mathbf{T}}^{T} \mathbf{S}_{f} {}^{t+\Delta t}\mathbf{f} + \hat{\mathbf{H}}{}^{t+\Delta t} \hat{\mathbf{r}} \right]-{}^{t}\hat{\mathbf{d}} \right)-\left( \frac{\delta}{\beta}-1 \right){}^{t}\dot{\hat{\mathbf{d}}}+\left(\Delta t -\frac{\delta \Delta t}{2\beta} \right) {}^{t}\ddot{\hat{\mathbf{d}}},\\
        {}^{t+\Delta t}\ddot{\hat{\mathbf{d}}} &= \frac{1}{\beta \Delta t^{2}} \left( \left[ \hat{\mathbf{H}}\hat{\mathbf{T}}^{T}\mathbf{S}_{f}{}^{t+\Delta t} {\mathbf{f}}+{\hat{\mathbf{H}}} {}^{t+\Delta t}\hat{\mathbf{r}}\right]-{}^{t}\hat{\mathbf{d}} \right)-\frac{1}{\beta \Delta t}{}^{t}\dot{\hat{\mathbf{d}}}-(\frac{1}{2 \beta}-1){}^{t}\ddot{\hat{\mathbf{d}}} .
    \end{align}
    \end{subequations}
    
Furthermore, Eq.~\eqref{eq:Selection_red} could be redefined using Eqs.~\eqref{eq:response_numerical_total} and \eqref{eq:response_numerical_derivatives} as

\begin{subequations}\label{eq:response_reorganized}
\begin{align}
\mathbf{z}_{n}\approx
\hat{\mathbf{S}}\mathbf{G}{}^{t+\Delta t}\mathbf{f}+\hat{\mathbf{S}}{}^{t+\Delta t}\mathbf{g},\\
\mathbf{G}=\left[ \begin{array}{c} \hat{\mathbf{H}}\hat{\mathbf{T}}^{T}\mathbf{S}_{f}\\
\frac{\delta}{\beta \Delta t}\hat{\mathbf{H}}\hat{\mathbf{T}}^{T}\mathbf{S}_{f}\\
\frac{1}{\beta {\Delta t}^{2}}\hat{\mathbf{H}}\hat{\mathbf{T}}^{T}\mathbf{S}_{f}
\end{array} \right],~
{}^{t+\Delta t}\mathbf{g}=\left[ \begin{array}{c}
\hat{\mathbf{H}}{}^{t+\Delta t}\hat{\mathbf{r}}\\
\frac{\delta}{\beta \Delta t}\left(\hat{\mathbf{H}}{}^{t+\Delta t}\hat{\mathbf{r}} -{}^{t}\hat{\mathbf{d}}\right)-\left( \frac{\delta}{\beta}-1 \right){}^{t}\dot{\hat{\mathbf{d}}}+\left(\Delta t -\frac{\delta \Delta t}{2\beta} \right) {}^{t}\ddot{\hat{\mathbf{d}}}\\
\frac{1}{\beta {\Delta t}^{2}}\left(\hat{\mathbf{H}}{}^{t+\Delta t}\hat{\mathbf{r}} -{}^{t}\hat{\mathbf{d}}\right)-\frac{1}{\beta \Delta t}{}^{t}\dot{\hat{\mathbf{d}}}-(\frac{1}{2 \beta}-1){}^{t}\ddot{\hat{\mathbf{d}}}
\end{array} \right].
\end{align}
\end{subequations}

    By substituting Eq.~\eqref{eq:response_reorganized} into Eq.~\eqref{eq:error}, the damped least-squares error equation can be rewritten as
    \begin{equation}\label{eq:plugged_error}
        \mathbf{J}(\mathbf{f},\alpha)=\left\|{}^{t+\Delta t}{\mathbf{z}}_{m}-\left(\hat{\mathbf{S}}\mathbf{G}{}^{t+\Delta t}\mathbf{f}+\hat{\mathbf{S}}{}^{t+\Delta t}\mathbf{g}\right)\right\|^{2}+\alpha\mathbf{I}\left\|{}^{t+\Delta t} {\mathbf{f}}\right\|^{2}.
    \end{equation}
    
    To minimize the damped least-squares error, the derivative of Eq.~\eqref{eq:plugged_error} with respect to the applied force can be computed as
    
    \begin{equation}\label{eq:error_diff}
        \frac{\partial\mathbf{J}}{\partial\mathbf{f}}=-2\mathbf{G}^{T}\hat{\mathbf{S}}^{T}\left({}^{t+\Delta t}{\mathbf{z}}_{m}-\left(\hat{\mathbf{S}}\mathbf{G}{}^{t+\Delta t}\mathbf{f}+\hat{\mathbf{S}}{}^{t+\Delta t}\mathbf{g}\right)\right)+2\alpha\mathbf{I} \left({}^{t+\Delta t} {\mathbf{f}} \right).
    \end{equation}
    
    Assuming that the solution to Eq.~\eqref{eq:error_diff} is zero and solving Eq.~\eqref{eq:error_diff} with respect to the applied force vector, the identified optimal force vector can be derived as

    \begin{equation}\label{eq:forceid_disp}
        {}^{t+\Delta t} {\mathbf{f}}=\left(\mathbf{G}^{T}\hat{\mathbf{S}}^{T}\hat{\mathbf{S}}\mathbf{G} + \alpha\mathbf{I} \right)^{-1}\mathbf{G}^{T}\hat{\mathbf{S}}^{T} \left({}^{t+\Delta t}{\mathbf{z}}_{m}-\hat{\mathbf{S}}{}^{t+\Delta t}\mathbf{g} \right).
    \end{equation}
    
    This is the final form of the force identification equation based on the reduced FE model and Tikhonov regularization method.
    Note that the force identification equation is derived using the conventional Newmark-$\beta$ method. Therefore, the response vectors of the reduced-order model can be computed using Eq.~\eqref{eq:response_reorganized}

    \begin{equation}\label{eq:response_compute}
    \left[\begin{array}{c} {}^{t+\Delta t}\hat{\mathbf{d}}\\{}^{t+\Delta t}\dot{\hat{\mathbf{d}}}\\{}^{t+\Delta t}\ddot{\hat{\mathbf{d}}} \end{array}\right]=\mathbf{G}{}^{t+\Delta t}\mathbf{f}+{}^{t+\Delta t}\mathbf{g}.
    \end{equation}
    
   As previously mentioned, the computed response vector (${}^{t+\Delta t}\hat{\mathbf{d}}$) is the reduced response vector on the reduced generalized coordinates. Therefore, to obtain physical responses, such as displacement, pressure, and acceleration, domain transformation of the response vectors is required. The domain transformation of the response vectors can be expressed as
   
   \begin{equation}\label{eq:response_recover}
        {}^{t+\Delta t}\mathbf{d}=\hat{\mathbf{T}}{}^{t+\Delta t}\hat{\mathbf{d}},~
    {}^{t+\Delta t}\dot{\mathbf{d}}=\hat{\mathbf{T}}{}^{t+\Delta t}\dot{\hat{\mathbf{d}}},~
    {}^{t+\Delta t}\ddot{\mathbf{d}}=\hat{\mathbf{T}}{}^{t+\Delta t}\ddot{\hat{\mathbf{d}}}.
   \end{equation}
   
   By implementing the proposed sequential computation process iteratively, the unmeasured applied force and status of the target structure can be identified. This enables the precise real-time identification of multiple responses for the entire target structure domain, including displacements, accelerations, strains, fluid pressure, and any other responses that can be computed by conventional FE dynamic analysis. The entire iterative identification process is represented by the reorganized and simplified forms in Table.~\ref{table:integration}.
%
%
\begin{table}[H]
    \centering
    \caption{Status identification process}
    \begin{tabular}{l}
        \hline
        1. Initialization. \\
        \hline
        1-1. Initialize response vectors\\
        \hspace{3mm}${}^{0}\hat{\mathbf{d}}, {}^{0}\dot{\hat{\mathbf{d}}}, {}^{0}\ddot{\hat{\mathbf{d}}}$\\
        1-2. Determine identification coefficients\\
        \hspace{3mm}$\beta, \delta, \alpha$\\
        \hline
        2. Applied force identification.\\
        \hline
        2-1. Compute internal force term\\
        \hspace{3mm}${}^{t+\Delta t}\hat{\mathbf{r}}=\hat{\mathbf{A}}\left[ (\frac{1}{\beta \Delta t^2}){}^{t}\hat{\mathbf{d}} + (\frac{1}{\beta \Delta t}){}^{t}\dot{\hat{\mathbf{d}}} + (\frac{1}{2 \beta}-1){}^{t}\ddot{\hat{\mathbf{d}}}\right] +\hat{\mathbf{D}}\left [(\frac{\delta}{\beta \Delta t}){}^{t}\hat{\mathbf{d}} + (\frac{\delta}{\beta}-1){}^{t}\dot{\hat{\mathbf{d}}}+(\frac{\delta \Delta t}{2\beta}-\Delta t){}^{t}\ddot{\hat{\mathbf{d}}}\right]$\\
        \hspace {3mm}${}^{t+\Delta t}\mathbf{g}=\left[ \begin{array}{c}
        \hat{\mathbf{H}}{}^{t+\Delta t}\hat{\mathbf{r}}\\
        \frac{\delta}{\beta \Delta t}\left(\hat{\mathbf{H}}{}^{t+\Delta t}\hat{\mathbf{r}} -{}^{t}\hat{\mathbf{d}}\right)-\left( \frac{\delta}{\beta}-1 \right){}^{t}\dot{\hat{\mathbf{d}}}+\left(\Delta t -\frac{\delta \Delta t}{2\beta} \right) {}^{t}\ddot{\hat{\mathbf{d}}}\\
        \frac{1}{\beta {\Delta t}^{2}}\left(\hat{\mathbf{H}}{}^{t+\Delta t}\hat{\mathbf{r}} -{}^{t}\hat{\mathbf{d}}\right)-\frac{1}{\beta \Delta t}{}^{t}\dot{\hat{\mathbf{d}}}-(\frac{1}{2 \beta}-1){}^{t}\ddot{\hat{\mathbf{d}}}
        \end{array} \right]$\\
        2-2. Identify applied force\\
        \hspace{3mm}${}^{t+\Delta t} {\mathbf{f}}=\left(\mathbf{G}^{T}\hat{\mathbf{S}}^{T}\hat{\mathbf{S}}\mathbf{G} + \alpha\mathbf{I} \right)^{-1}\mathbf{G}^{T}\hat{\mathbf{S}}^{T}({}^{t+\Delta t}{\mathbf{z}}_{m}-\hat{\mathbf{S}}{}^{t+\Delta t}\mathbf{g})$\\
        \hline
        3. Calculate state vectors\\
        \hline
        \hspace{3mm}$\left[\begin{array}{c} {}^{t+\Delta t}\hat{\mathbf{d}}\\{}^{t+\Delta t}\dot{\hat{\mathbf{d}}}\\{}^{t+\Delta t}\ddot{\hat{\mathbf{d}}} \end{array}\right]=\mathbf{G}{}^{t+\Delta t}\mathbf{f}+{}^{t+\Delta t}\mathbf{g}$\\
        \hline
    \end{tabular}
    \label{table:integration}
\end{table}

\section{Numerical test} \label{section3}
\begin{figure}[H]
    \centering
    \includegraphics[width=1.0\textwidth]{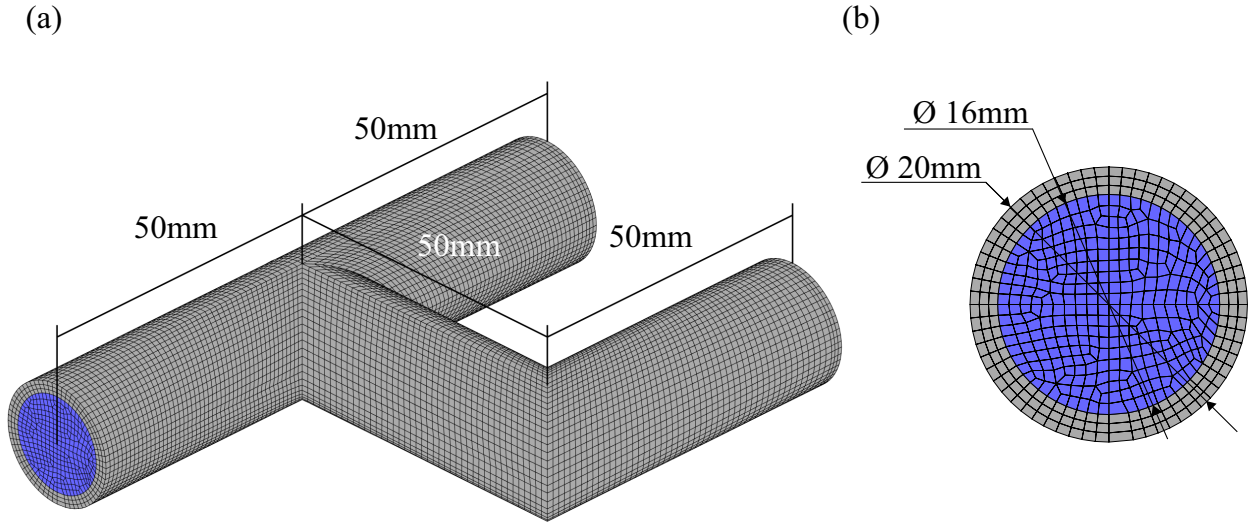}
    \captionof{figure}{Geometry of numerical test structure}
\label{fig:Numeric_setup}
\end{figure}

The accuracy and efficiency of the proposed algorithm are first evaluated using a simple numerical testbed. An h-shaped pipeline structure filled with water is chosen as a target structure. Figure.~\ref{fig:Numeric_setup} shows the geometry of the target structure and the generated FE mesh.

\begin{figure}[H]
    \centering
    \includegraphics[width=0.6\textwidth]{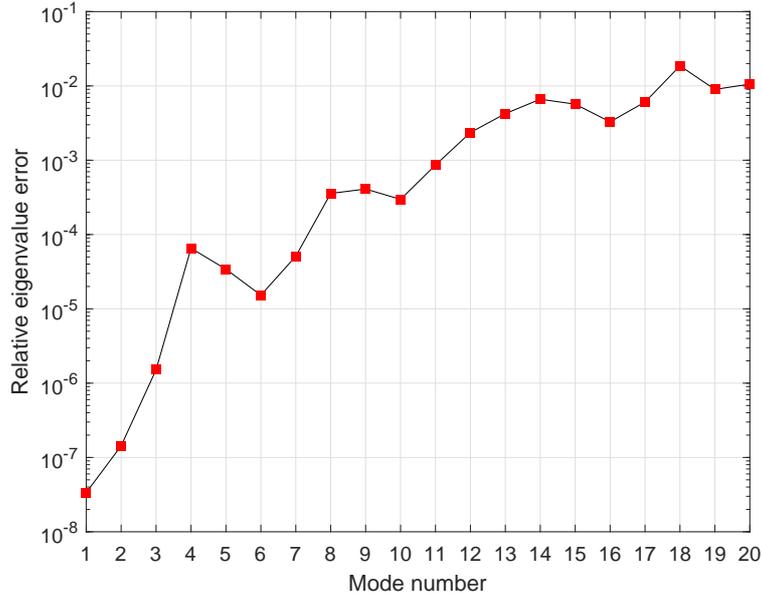}
    \captionof{figure}{Relative eigenvalue error of reduced model}    
\label{fig:Reduction_error_numeric}
\end{figure}

The Young modulus ($E$) of the structure is 210,000 $\si{\giga \pascal}$, and the Poisson ratio ($\nu$) is set to 0.3. The density of the structural part ($\rho^{s}$) is set to 8.0E-09 $\si{\tonne}$/$\si{\milli \meter}{}^{3}$. The values of the speed of sound in water $c$ and density of the fluid part ($\rho^{f}$) are adopt 1480 $\si{\meter}$/$\sec$ and 1.01E-09 $\si{\tonne}$/$\si{\milli \meter}{}^{3}$, respectively. 
A strongly coupled vibroacoustic model order reduction is then performed on the generated FE model for efficiency. The dominant mode counts of the structure ($N_{d}^{s}$) and fluid ($N_{d}^{f}$) domains are both set to 30 modes.
After the reduction process, the entire DOF count of the system is reduced from 224742 to 60.
Figure.~\ref{fig:Reduction_error_numeric} shows the relative eigenvalue errors between the full and reduced models.

The results show that the reduced model can preserve the vibroacoustic properties of the original full model, even if it is reduced to an efficient form.

\subsection{Numerical test condition}
In the numerical study, the following aspects are considered and evaluated: 1) the accuracy of the identified values, 2) stability under noisy excitation conditions, and 3) computational efficiency. The numerical solution of the test problem is pre-computed before performing the force identification stage, and the acceleration signals of the very limited measurement points are then collected to simulate the \textit{"experimentally measured"} input signals. The proposed force identification algorithm is then implemented to identify the applied forces and responses of the target system using only the collected \textit{"experimentally measured"} signals. The accuracy of the proposed algorithm is evaluated by comparing the \textit{"identified"} quantities to the \textit{"experimentally measured"} signals. 

\begin{figure}[H]
    \centering
    \includegraphics[width=0.7\textwidth]{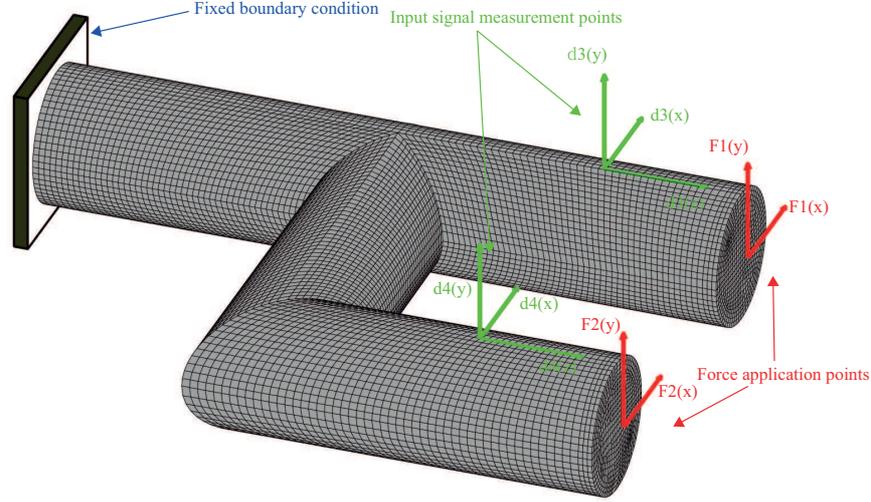}
    \captionof{figure}{Geometry of numerical test structure}    
\label{fig:Numeric_BC}
\end{figure}

One side of the target structure is constrained by the fixed boundary condition, and external forces are applied on the other sides of the structure in two orthogonal directions (denoted as x and y in Fig~\ref{fig:Numeric_BC}). The \textit{"experimentally measured"} acceleration signals are then collected from points on the pipe surface 20 $\si{\milli \meter}$ from the force application points. Figure.~\ref{fig:Numeric_BC} shows the applied boundary conditions and the response measurement points.

The applied external force profiles are set to combinations of multiple arbitrary sinusoidal functions as follows:

\begin{subequations}
    \begin{align}
        \mathbf{f}^{1}_{x}(t)&=200sin(30\pi t)+370sin(175\pi t),\\
        \mathbf{f}^{1}_{y}(t)&=500sin(100\pi t)+460sin(95\pi t),\\
        \mathbf{f}^{2}_{x}(t)&=460sin(150\pi t)+280sin(30\pi t),\\
        \mathbf{f}^{2}_{y}(t)&=280sin(120\pi t)+370sin(23\pi t),
    \end{align}
\end{subequations}
where subscripts $x$ and $y$ are the force application and measurement directions, respectively.
    
    Pre-computation of the numerical solution is performed using the conventional Newmark-$\beta$ time integration method, and 10 $\si{\sec}$ of numerical solutions are computed to generate the reference signal for the numerical test. The \textit{"experimentally measured"} acceleration signals are collected from the measurement points and used as the input signal for the numerical test. In practice, directly measured sensor signals may suffer from measurement/numerical errors for various reasons. Hence, to simulate the practically measured sensor signals, the collected acceleration signals are polluted by adding random Gaussian noise as
    
\begin{equation}
    \ddot{\bar{\mathbf{d}}}=\ddot{{\mathbf{d}}}+ \tau \sigma \mathbf{N},
\end{equation}
where $\tau$ and $\sigma$ are the rate coefficients of the noise and the standard deviation of the collected acceleration signals, respectively, and $\mathbf{N}$ is the standard normal distribution function. $\ddot{\bar{\mathbf{d}}}$ is polluted response vector. The distribution value of the noise is set to 1 $\%$ ($\tau$=0.01) of the distribution of the measured acceleration signal. 
The polluted acceleration signals are then used as input signals, as

\begin{equation}
    {{\mathbf{z}}}_{m}=\left[ \begin{array}{c}
    \ddot{\bar{\mathbf{d}}}^{3}_{x}\\
    \ddot{\bar{\mathbf{d}}}^{3}_{y}\\
    \ddot{\bar{\mathbf{d}}}^{3}_{z}\\
    \ddot{\bar{\mathbf{d}}}^{4}_{x}\\
    \ddot{\bar{\mathbf{d}}}^{4}_{y}\\
    \ddot{\bar{\mathbf{d}}}^{4}_{z}
    \end{array} \right].
\end{equation}

The desired applied forces and response values ($\mathbf{f}_{x,y}^{1,2},\mathbf{d}_{x,y,z}^{1,2}$) are identified using the collected signals as the input data for the proposed force identification process.
The sampling rate of the numerical evaluation test is set to 1000 Samples $/\si{\sec}$, and hence, the identification time increment value $\Delta t$ is 0.001 $\si{\sec}$.
In addition, the required computation time for the identification process is investigated to evaluate the efficiency of the proposed identification algorithm.

The identified quantities are time-transient results. Hence, proper correlation measures are required to evaluate the historical agreement between the identified and their reference results. In this study, Geer’s error measures\cite{geers1984objective,whang1994two} are implemented to quantify the errors in the identified quantities. The errors between the identified and reference signals are quantified using Geer’s following three error measures:
    \begin{equation}
        \epsilon_{mag}=\frac{\sqrt{\Sigma{(\mathbf{z}_{ni}^2)}}}{\sqrt{\Sigma{(\mathbf{z}_{mi}^2)}}}-1.0,~\epsilon_{phase}=1.0-\frac{\sqrt{|\Sigma{(\mathbf{z}_{ni}\mathbf{z}_{mi})}|}}{\sqrt{\sqrt{\Sigma{(\mathbf{z}_{ni}^{2})}}\sqrt{\Sigma{(\mathbf{z}_{mi}^{2})}}}}~,\epsilon_{comp}=\sqrt{\epsilon_{mag}^{2}+\epsilon_{phase}^{2}},
        \end{equation}
where the subscript $i$ denotes the $i$th element of the identified result vector. The magnitude error $\epsilon_{mag}$ is the measure that can quantify the magnitude differences between the identified and measured results. The magnitude of the error measure can exceed 1. The second measure $\epsilon_{phase}$ can quantify the phase difference between identified and measured signal vectors. This value does not exceed 1, and bounded between 0 to 1.
The third measure $\epsilon_{comp}$ is a comprehensive error measure that considers both the magnitude and phase errors of the identified quantities. This value can also exceed 1. For all these error quantification measures, a smaller value indicates good agreement between the compared quantities.

\subsection{Numerical test results}
\begin{figure}[H]
    \centering
    \includegraphics[width=0.9\textwidth]{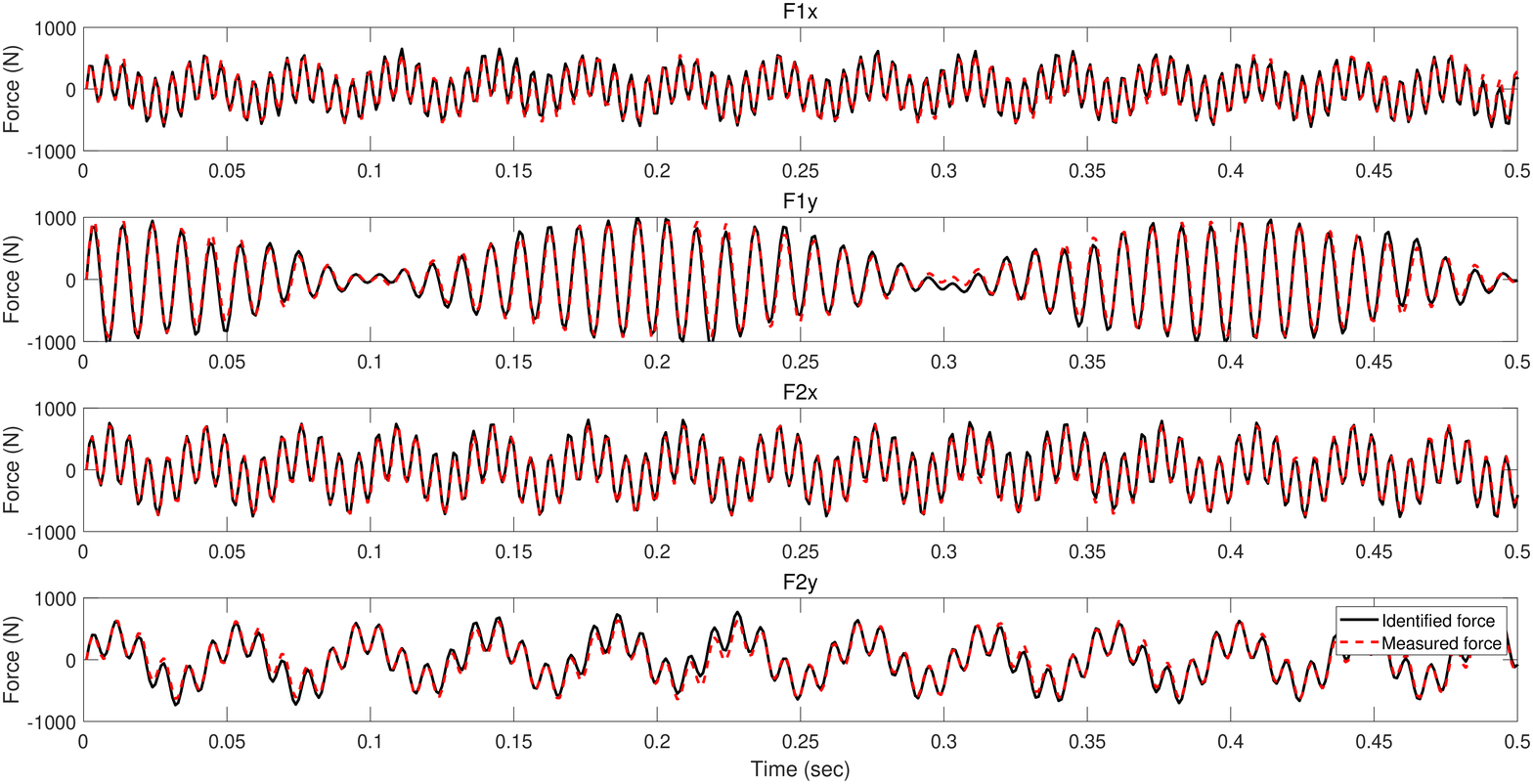}
    \captionof{figure}{Identified force results}    
\label{fig:Identified_force_numeric}
\end{figure}

Through the numerical test, the applied external force and unknown displacement data are identified and evaluated by comparing the results to the pre-computed reference information.
Figure.~\ref{fig:Identified_force_numeric} shows the identified force results.

The results indicate that the identified external forces can describe the applied force signals. The three error measurement values for the identified results are listed in Table~\ref{table:Error_measure_identified_force}.
\begin{table}[H]
    \centering
    \caption{Error measure values of identified forces}
    \begin{tabular}{c|c|c|c|c}
    \hline
    &\multicolumn{4}{c}{Force direction}\\
    \hline
    Error measure&$\mathbf{f}^{1}_{x}$ & $\mathbf{f}^{1}_{y}$ &	$\mathbf{f}^{2}_{x}$ &	$\mathbf{f}^{2}_{y}$ \\
    \hline
    $\epsilon_{mag}$ & 1.755E-02 &	2.1426E-02 &	1.629E-02 &	4.493E-02 \\
    $\epsilon_{phase}$  & 7.280E-03 &	6.828E-03 &	4.412E-03 &	6.377E-03 \\
    $\epsilon_{comp}$  & 1.900E-02 &	2.248E-02 &	1.688E-02 &	4.538E-02 \\
    \hline
    \end{tabular}
    \label{table:Error_measure_identified_force}
\end{table}
All the error measure values ($\epsilon_{mag}$, $\epsilon_{phase}$, $\epsilon_{comp}$) of the identified forces are bounded under 0.04538. The results show that the identified results accurately describe the reference signals.

\begin{figure}[H]
    \centering
    \includegraphics[width=0.9\textwidth]{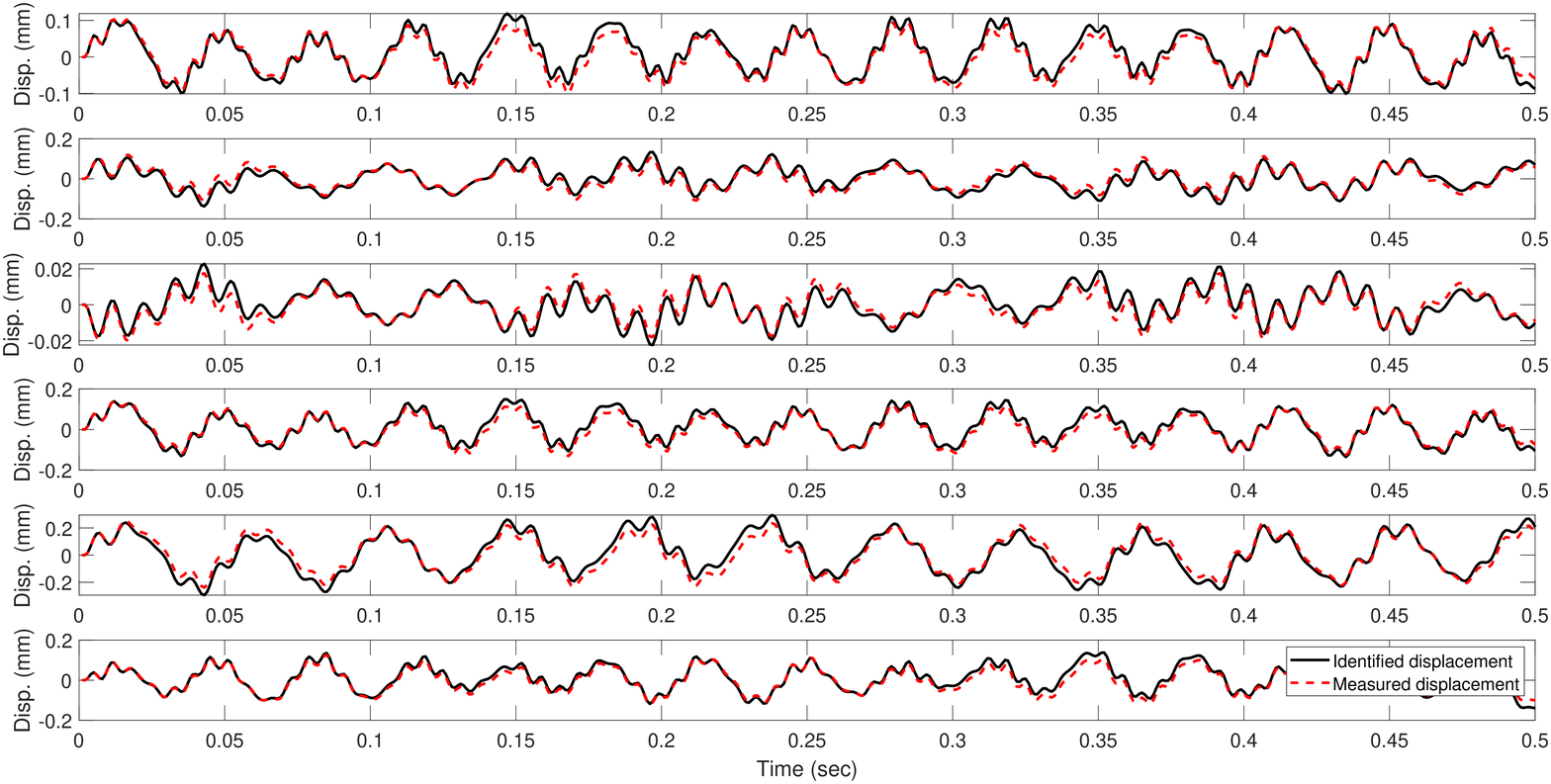}
    \captionof{figure}{Identified displacement results}    
\label{fig:Identified_displacement_numeric}
\end{figure}

The unmeasured displacement identification results are presented in Fig.~\ref{fig:Identified_displacement_numeric}.

The identified displacement results can also precisely describe the directly computed reference displacements. The identification errors of the displacement signals are also evaluated by applying Geer’s error factors. Table~\ref{table:Error_measure_identified_displacement} presents the computed error factors for the identified displacement signals.
\begin{table}[H]
    \centering
    \caption{Error measure values of identified displacements}
    \begin{tabular}{c|c|c|c|c|c|c}
    \hline
    &\multicolumn{6}{c}{Displacement direction}\\
    \hline
    Error measure&$\mathbf{d}^{1}_{x}$ & $\mathbf{d}^{1}_{y}$ &	$\mathbf{d}^{1}_{z}$ &	$\mathbf{d}^{2}_{x}$ &	$\mathbf{d}^{2}_{y}$ &	$\mathbf{d}^{2}_{z}$\\
    \hline
    $\epsilon_{mag}$ & 7.361-02 &	5.527E-02 &	5.940E-02 &	7.491E-02 &	6.415E-02 &	7.546E-02 \\
    $\epsilon_{phase}$  & 2.312E-02 &	2.134E-02 &	2.781E-02 &	2.271E-02 &	1.319E-02 &	2.369E-02 \\
    $\epsilon_{comp}$  & 7.715E-02 &	5.925E-02 &	6.559E-02 &	7.828E-02 &	6.549E-02 &	7.909E-02 \\
    \hline
    \end{tabular}
    \label{table:Error_measure_identified_displacement}
\end{table}

All error measures of the identified displacement signals are computed as values smaller than 0.079. Hence, the identified displacement results accurately describe the trend of the reference displacement signals.

The computational efficiency of the proposed algorithm is also evaluated by comparing the required computation time with the actual pre-computed numerical test duration.
The required computation time for the numerical test is 0.324939 $\si{\sec}$ for 10 $\si{\sec}$ of reference signals.
According to the efficiency test results, the process can be implemented on the test structure even in real time under the intended sample rates. The results demonstrate that the proposed identification process can be implemented in practical applications.

\subsection{Stability of identified forces under noisy conditions}

\begin{figure}[H]
    \centering
    \includegraphics[width=0.9\textwidth]{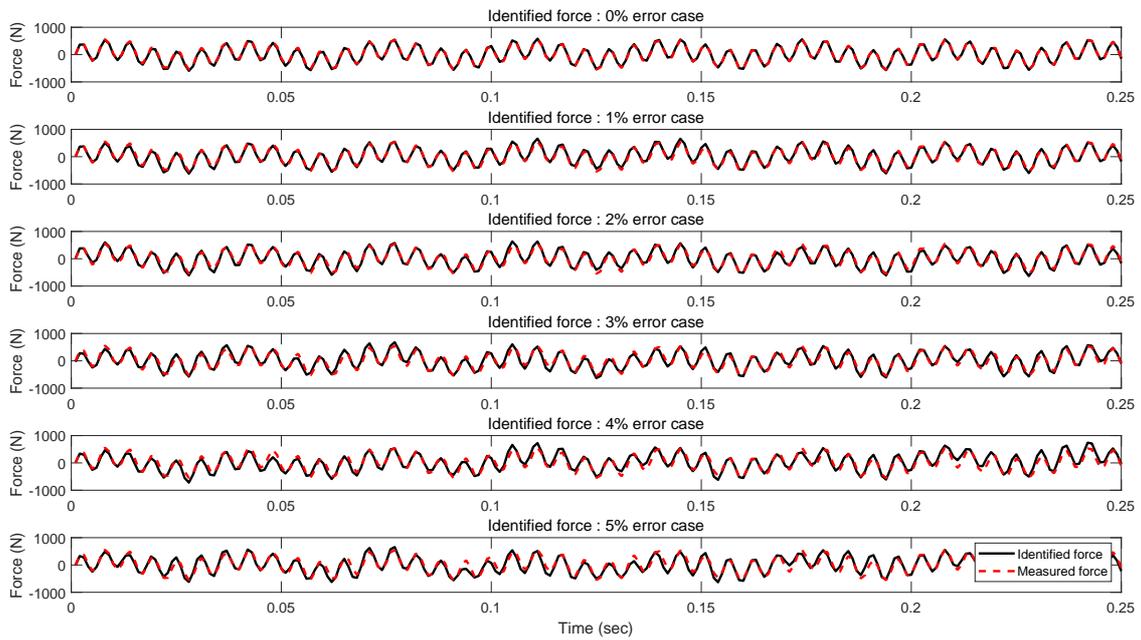}
    \captionof{figure}{identified force results of each error test case}    
\label{fig:force_1to5}
\end{figure}

\begin{figure}[H]
    \centering
    \includegraphics[width=0.4\textwidth]{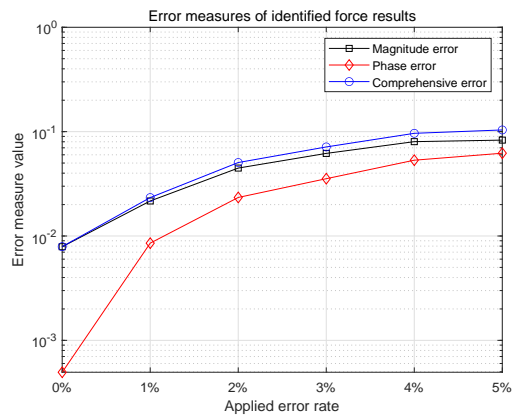}
    \captionof{figure}{Computed error measures of various error cases}    
\label{fig:force_1to5_plot}
\end{figure}

In practice, measured quantities such as acceleration and displacement can suffer from unintended noise. To test the stability of the force identification process under highly noisy conditions, the accuracy of the identified forces is evaluated under various noise levels. The test structure and other conditions are set to the same state as those in the previous numerical study. The noise rate coefficient $\tau$ is varied from 0 $\%$ to 5 $\%$ ($\tau$ = 0.00 to 0.05) in six increments. Each test case is performed 20 times iteratively, and three error coefficients are computed and averaged for all cases. The unknown applied forces are identified and their accuracy is evaluated. A single point of the identified force results ($\mathbf{f}_{x}^{1}$) is considered in the study, and the computed error levels of the remaining identified force results are observed to be similar to the represented results. The identified results are depicted in Fig~\ref{fig:force_1to5}. The computed error measures of the identified force data are also depicted in Fig~\ref{fig:force_1to5_plot}.

As shown in the figures, the identification errors gradually increase with the magnitude of the error. However, the error values are still bounded under 0.1314 for all cases and measures. The results clearly show that the proposed method can stably identify unmeasured applied forces even under noisy measurement conditions.

%
%
\subsection{Comparative test with conventional method}
To test the relative performance of the proposed method, a comparative test with the other methods is performed. The test is performed in the same system and conditions as in the previous numerical tests, and the results obtained from the proposed and AKF\cite{lourens2012augmented} methods are compared to the real applied force values. The time increments of the methods are set to different values by considering their numerical properties. In the case of the proposed method, the Newmark-$\beta$ time-integration algorithm employed is unconditionally stable. Hence, the proposed method can be performed with relatively large time increment values. In this study, the time increment for the proposed method is set to 1E-04 $\si{\sec}$. Otherwise, the solution to the AKF method is conditionally stable. Hence, sufficiently small time increments are required to implement the method. Therefore, various time increment values (from 2e-07 to 2E-08 $\si{\sec}$) are tested for the AKF algorithm to check its convergence and relative accuracy. For all the test cases, 0.1 $\si{\sec}$ of identified force results and their required computation times are collected. The noise rate coefficient is set to 1 $\%$ ($\tau = 0.01$) for all cases. In the original paper on the AKF method, the problem only considered monophysical structural systems without a reduced-order modeling process. Hence, the formulations of the method are modified by applying the strongly coupled vibroacoustic reduced-order modeling technique for equivalent comparison. A detailed derivation of the modified formulations can be found in \ref{appendix:B}. Figure.~\ref{fig:Comparison_study} shows the first 0.005 $\si{\sec}$ of identified force results of the comparative study cases.

\begin{figure}[H]
    \centering
    \includegraphics[width=1.0\textwidth]{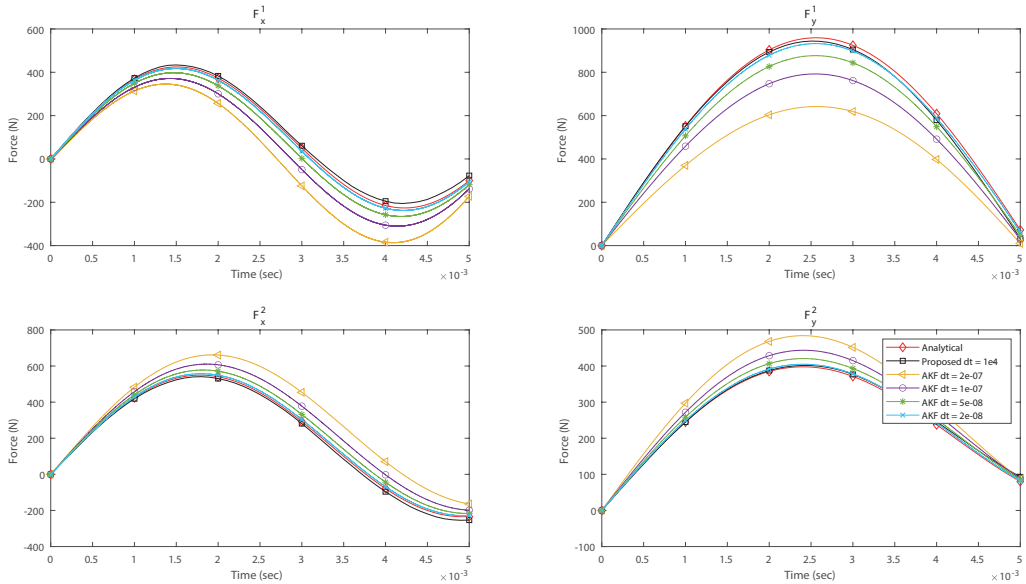}
    \captionof{figure}{Identified forces of test cases}
    \label{fig:Comparison_study}
    \end{figure}
    
\begin{figure}[H]
    \centering
    \includegraphics[width=1.0\textwidth]{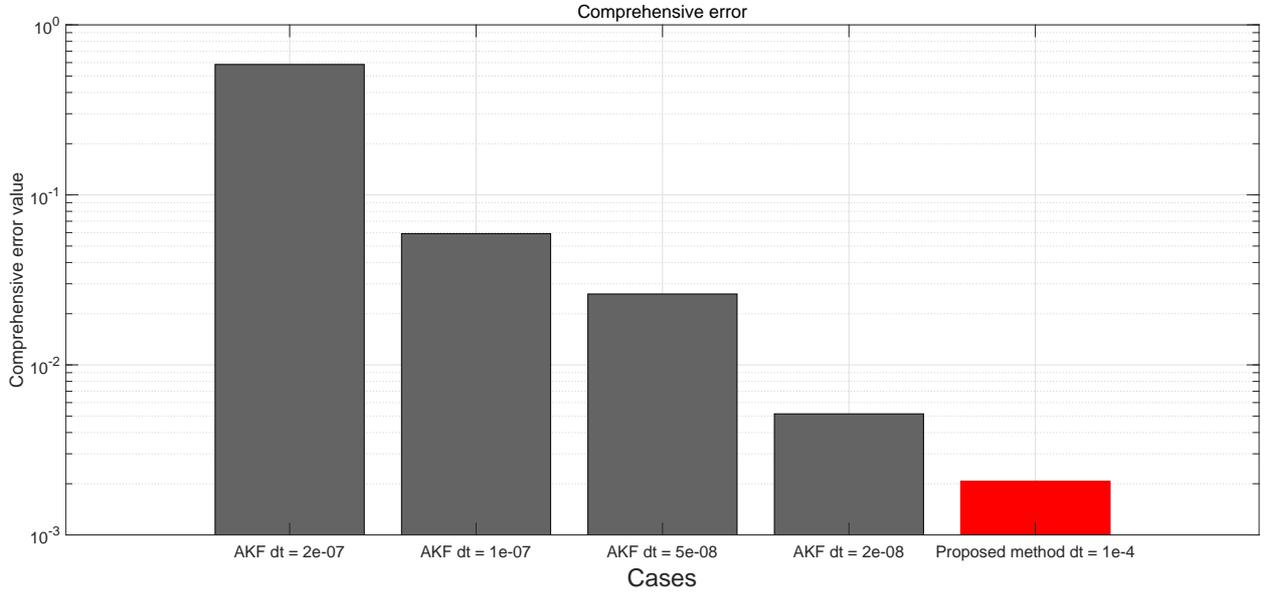}
    \captionof{figure}{Comprehensive error values of test cases}    
    \label{fig:Comparison_comprehensive}
    \end{figure}

The results reveal that the proposed algorithm can accurately identify unknown applied forces even with relatively large time increment values compared to the AKF method. Figure.~\ref{fig:Comparison_comprehensive} shows the comprehensive error values of force element $\mathbf{f}_{x}^{1}$ for each case.

The computed comprehensive errors of the results from the AKF method cases decreased by reducing the size of the time increment. Meanwhile, the proposed method with a time increment of 1E-04 $\si{\sec}$ could give more accurate results than the results from the AKF with a time increment of 2E-08 $\si{\sec}$. The required computation time for the test is listed in Table.~\ref{table:Comparative_test_etime}.

\begin{table}[H]
    \centering
    \caption{Required computation time for identification}
 \begin{tabular}{c|c|c}
\hline
Used method &Time increments &Required time for computation ($\si{\sec}$)\\
\hline
   Proposed method &1E-04 &0.124\\
\hline
& 2E-07 &116.8754\\
AKF& 1E-07 &250.4187\\
& 5E-08 &516.6280\\
& 2E-08 &997.7140\\
\hline
\end{tabular}
    \label{table:Comparative_test_etime}
\end{table}

The required computation time for the proposed method is less than that of the AKF method. The results reveal that the proposed algorithm can be implemented in vibroacoustic systems, even with larger time increment requirements.

%
%

\section{Experimental test}\label{section4}
The accuracy of the identified quantities and the computational efficiency of the proposed algorithm are also evaluated experimentally. A simple water-filled pipeline structure is constructed as the experimental target structure. A vibroacoustic FE model of the structure is generated, and the proposed identification algorithm is then implemented on a PC. An experimental evaluation is performed by comparing the identified results with the experimentally measured quantities. In addition, the computational efficiency is evaluated at the end of this section.

The target structure of the testbed is a simple pipeline with an elbow component bent at 90$^{\circ}$ in the middle of the structure. The pipe system is then fixed to a rigid fixture at both ends. The length of each straight section is 1m, and the diameters of the outer and inner sides of the pipeline are 27.2 $\si{\milli \meter}$ and 23.9 $\si{\milli \meter}$, respectively. The entire target pipe structure is made of AISI 304 stainless steel and filled with water.

\begin{figure}[H]
    \centering
    \includegraphics[width=0.9\textwidth]{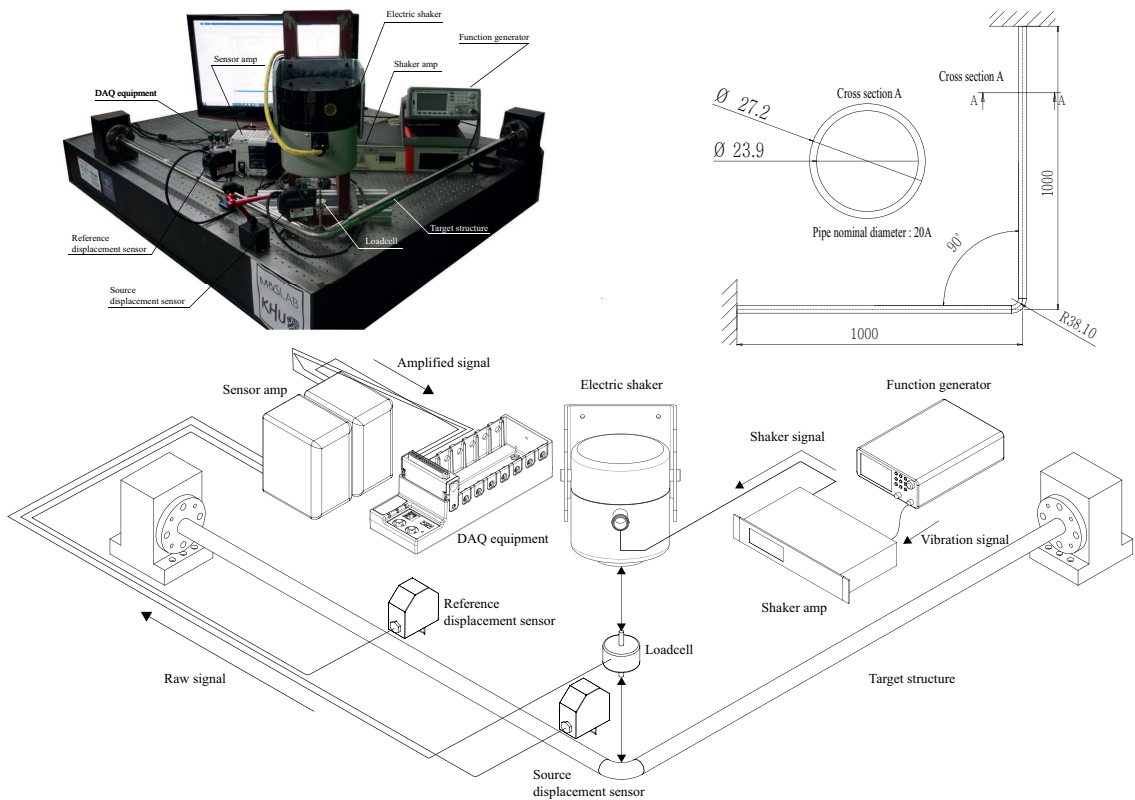}
    \captionof{figure}{Experimental setup}    
\label{fig:Experimental_setup}
\end{figure}

For the experimental study, the displacement is measured at two separate points. The first point is the source displacement, which is used as the measured displacement signal for the force identification process and is measured at the same point where the external force is applied. A laser displacement sensor (LK-G30, KEYENCE) is used to measure the source displacement signal of the structure. An additional laser displacement sensor (LK-G150) is mounted in the middle of the straight section of the structure and is used as the reference displacement signal to evaluate the identified result. Meanwhile, an external force is applied to the structure through a modal shaker (ET-139, LabWorks) on the top of the elbow component. The excitation signal is generated using a function generator (33500B, Keysight) and amplified using a signal amplifier (PA-138, LabWorks). The applied force is measured using a load cell (UMM 5 $\si{\kilo \gram}$f, DACELL). Figure.~\ref{fig:Experimental_setup} shows the constructed experimental testbed and the detailed geometry of the constructed target structure.

\begin{figure}[H]
    \centering
    \includegraphics[width=0.9\textwidth]{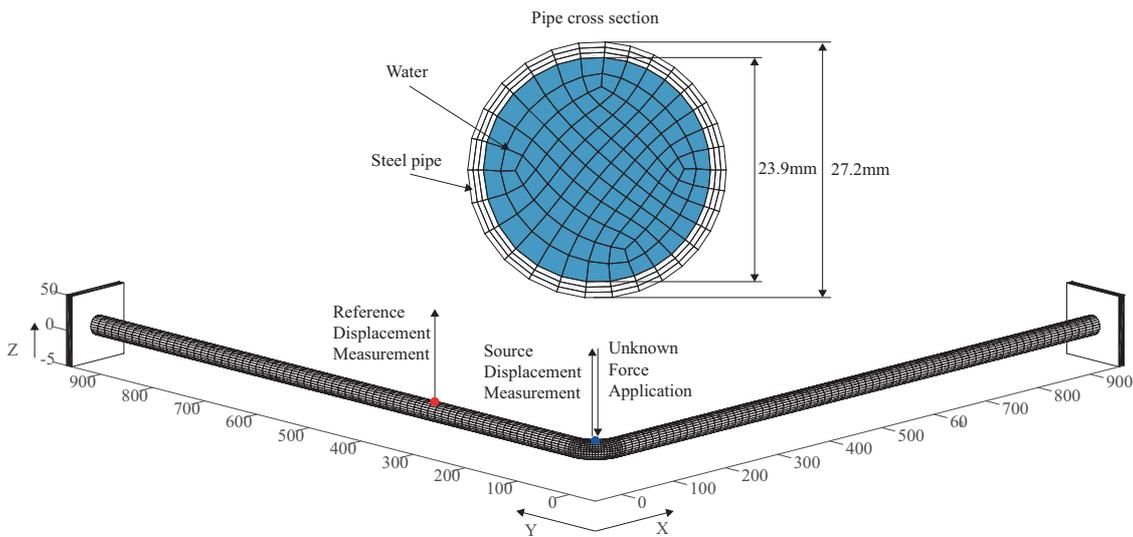}
    \captionof{figure}{Numerical setup}    
\label{fig:Mesh}
\end{figure}

The vibroacoustic FE model of the structure is generated in the same shape as the target structure using hexahedral elements. A total of 18 560 and 14 400 elements are used for the fluid and structure parts, respectively. Figure.~\ref{fig:Mesh} shows the FE model generated for the target structure.

The Young’s modulus $\mathbf{E}$ and density of the structure $\rho^{s}$ are 200 $\si{\giga \pascal}$ and 8E-09 $\si{\tonne}$/$\si{\milli \meter}^{3}$ respectively, and the speed of sound in water $c$ and the density of the fluid $\rho^{f}$ are 1480 $\si{\meter}$/$\si{\sec}$ and 1.01E-09 $\si{\tonne}$/$\si{\milli \meter}^{3}$, respectively. The Poisson ratio of the structural part is $0.3$. The material properties used are pre-calibrated values updated using conventional FE model updating techniques\cite{Rates_Fox, Friswell, ma1991sensitivity, dhandole2010comparative}.

\begin{figure}[H]
    \centering
    \includegraphics[width=0.5\textwidth]{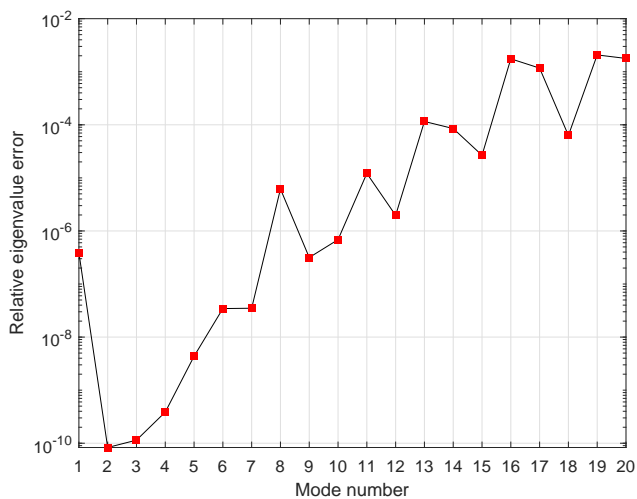}
    \captionof{figure}{Relative eigenvalue error of reduced model}
\label{fig:Reduction_error_exp}
\end{figure}

The generated FE model is then reduced to an efficient form using the strongly coupled vibroacoustic reduced-order modeling technique. The modes used for the reduction process are 30 for the structural part ($N_{d}^{s}$) and 30 for the fluid part ($N_{d}^{f}$) of the full model. Hence, the total DOF counts of the model is reduced from 79212 DOFs to 60 DOFs after reduction. Figure~\ref{fig:Reduction_error_exp} shows the relative eigenvalue errors between the original and reduced models.

\subsection{Experimental test conditions}
    Through the experimental study, the accuracy of the identified force and displacement results is first evaluated by comparing the results to the directly measured quantities under various harmonic excitation conditions. Random excitation conditions are also considered to investigate the stability of the identified forces under various noisy loading conditions. In addition, the computation time of the implemented identification algorithm is inspected to evaluate the computational efficiency of the proposed method. The computation time required to process the entire experimental measurement data set is compared with the actual measurement duration of the experiment. 
    For every test case, 30 $\si{\sec}$ of experimental data are acquired and processed, and the data acquisition sample rate is set to 10240 Samples/$\si{\sec}$. The identification frequency of the implemented algorithm is set to the same value.

    \subsection{Experimental test results}
    
    The accuracy of the force identified using the proposed method is evaluated by a real-time identification test under various sinusoidal excitation conditions. In this study, five different sinusoidal excitation conditions are used as test cases:1 $\si{Hz}$, 2 $\si{Hz}$, 4 $\si{Hz}$, 8 $\si{Hz}$, and 16 $\si{Hz}$. A single displacement signal is measured at the same position as the force application that is directly connected to the implemented software. The applied force is also directly measured at the force application point.
    
\begin{figure}[H]
    \centering
    \includegraphics[width=1.0\textwidth]{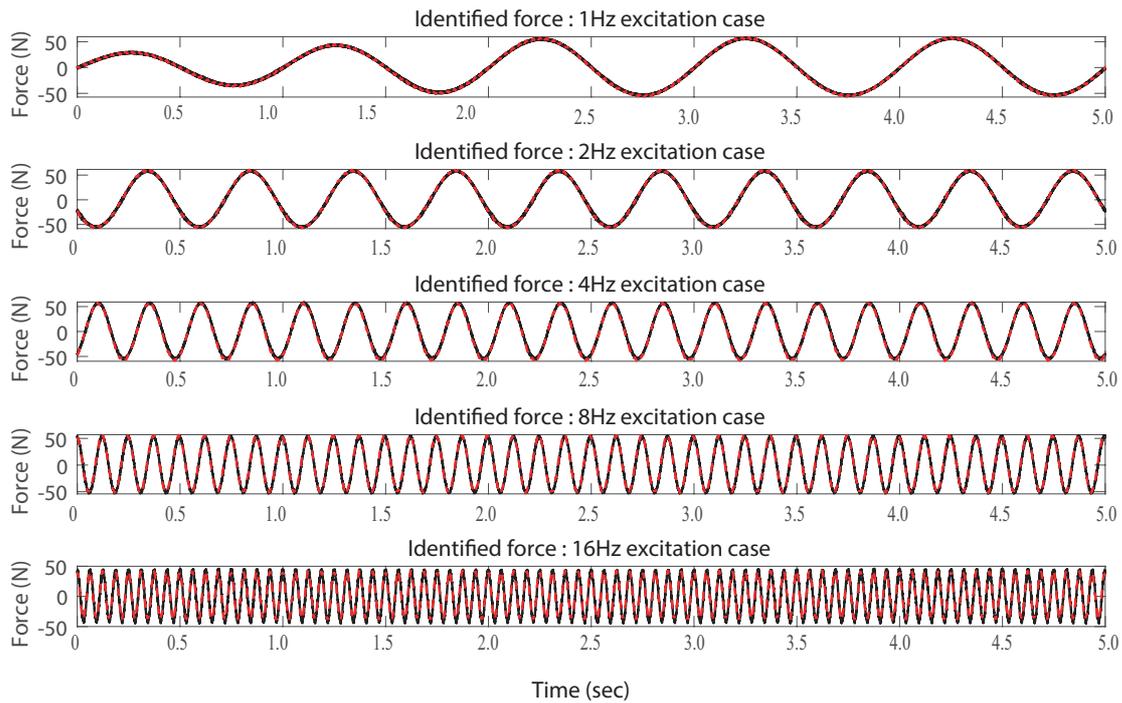}
    \captionof{figure}{Identified applied force under sinusoidal loading condition.}
    
\label{fig:Force_sinusoidal}
\end{figure}

\begin{figure}[H]
    \centering
    \includegraphics[width=0.5\textwidth]{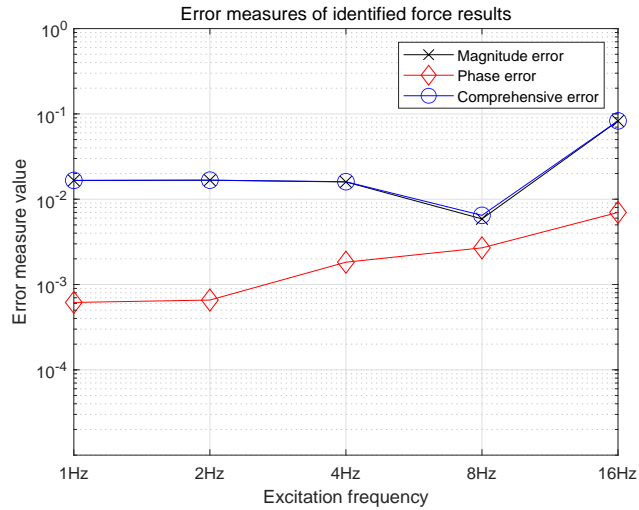}
    \captionof{figure}{Computed error measures for each identified force result}
\label{fig:error_sinu_force_exp}
\end{figure}

    The force identified through the software is then compared with the directly measured force data. Figure.~\ref{fig:Force_sinusoidal} shows the identified and directly measured applied forces for each case, and Fig.~\ref{fig:error_sinu_force_exp} shows their computed error measures.

    The measured applied force results are depicted by the dashed red line, and the identified force results are plotted as a solid black line. The results show that the force identified by the proposed method can precisely describe the actual applied force for all test cases. The error measure values are computed to be under 1.1333E-02 for all cases.
    
\begin{figure}[H]
    \centering
    \includegraphics[width=1.0\textwidth]{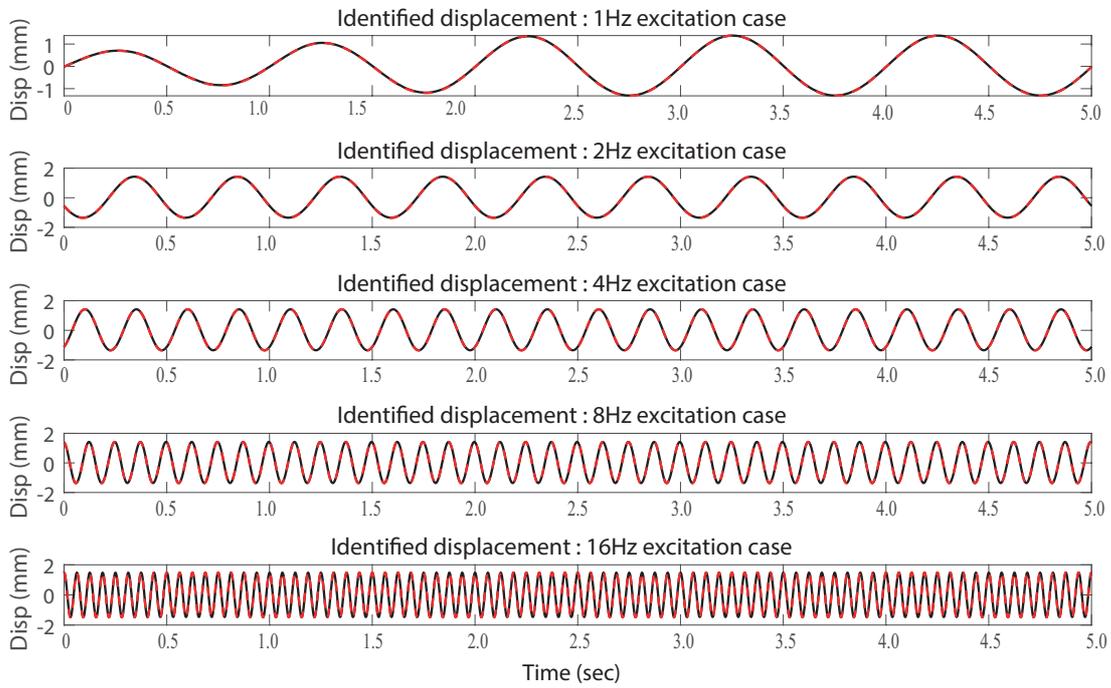}
    \captionof{figure}{Identified displacement response under sinusoidal loading condition.}    
    \label{fig:Displacement_sinusoidal}
    \end{figure}

\begin{figure}[H]
    \centering
    \includegraphics[width=0.5\textwidth]{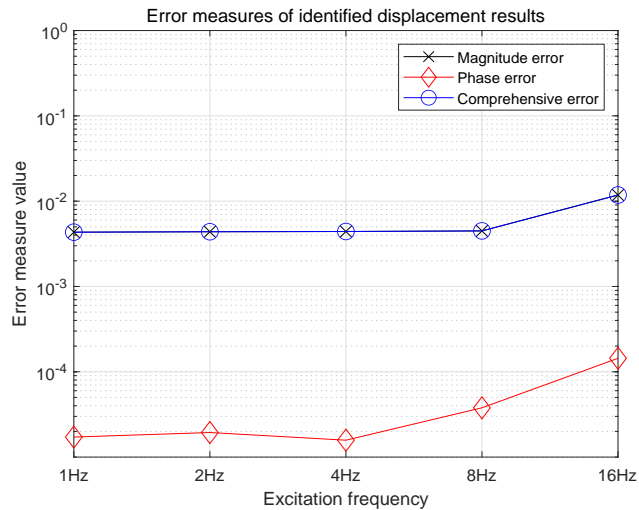}
    \captionof{figure}{Computed error measures for each identified displacement result}
\label{fig:error_sinu_disp_exp}
\end{figure}

    While performing the suggested force identification process, $\ddot{{\mathbf{d}}},\dot{{\mathbf{d}}},{\mathbf{d}}$ are also updated concurrently. The accuracy of the identified displacements is also evaluated by comparing them to the directly measured quantities. Figure~\ref{fig:Displacement_sinusoidal} shows the identified results, and their error measures are plotted in Fig~\ref{fig:error_sinu_disp_exp}.
    
    All identified results follow the directly measured quantities well, and the error measures are bounded under 6.935E-02. Hence, the displacements are successively identified.
    
\begin{figure}[H]
    \centering
    \includegraphics[width=1.0\textwidth]{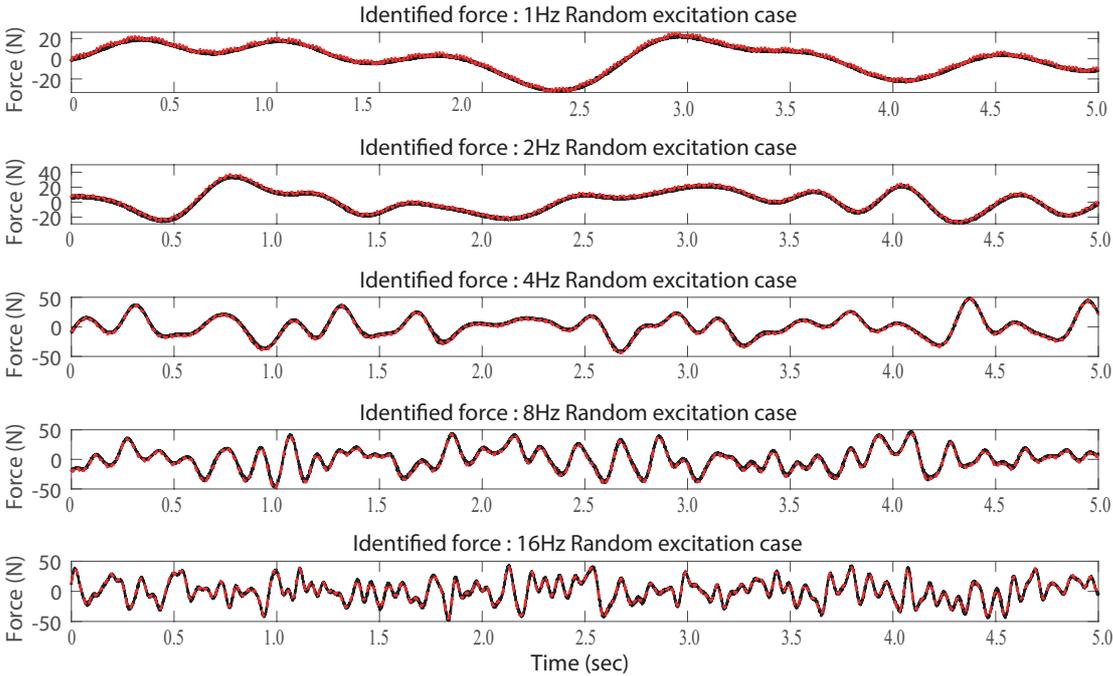}
    \captionof{figure}{Identified applied force under random loading condition.}
\label{fig:Force_Random}
\end{figure}

\begin{figure}[H]
    \centering
    \includegraphics[width=0.5\textwidth]{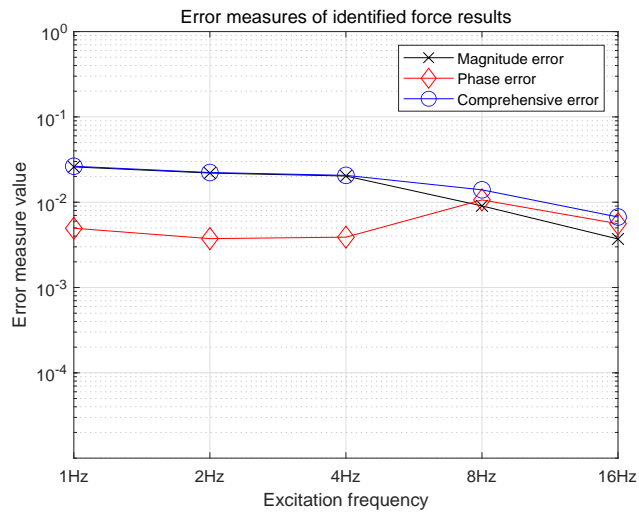}
    \captionof{figure}{Computed error measures for each identified force result}
\label{fig:error_random_force_exp}
\end{figure}

    The stability of the identified forces under various random excitation signals is also evaluated. For each test case, the excitation signal is generated as a composition of random sinusoidal signals in different frequency ranges. The frequency ranges used for the test cases are 1 $\si{Hz}$, 2 $\si{Hz}$, 4 $\si{Hz}$, 8 $\si{Hz}$, and 16 $\si{Hz}$. Figure.~\ref{fig:Force_Random} shows the results of the identification tests under random excitation conditions, and the conventions of the plotted lines are the same as those of the previous results. Figure~\ref{fig:error_random_force_exp} lists their computed error measures.
    
    The results demonstrate that the proposed method can accurately identify the applied force even under random excitation conditions. The error measures are also computed as smaller values than in the sinusoidal cases and are bounded under 1.963E-03.
    
\begin{figure}[H]
    \centering
    \includegraphics[width=1.0\textwidth]{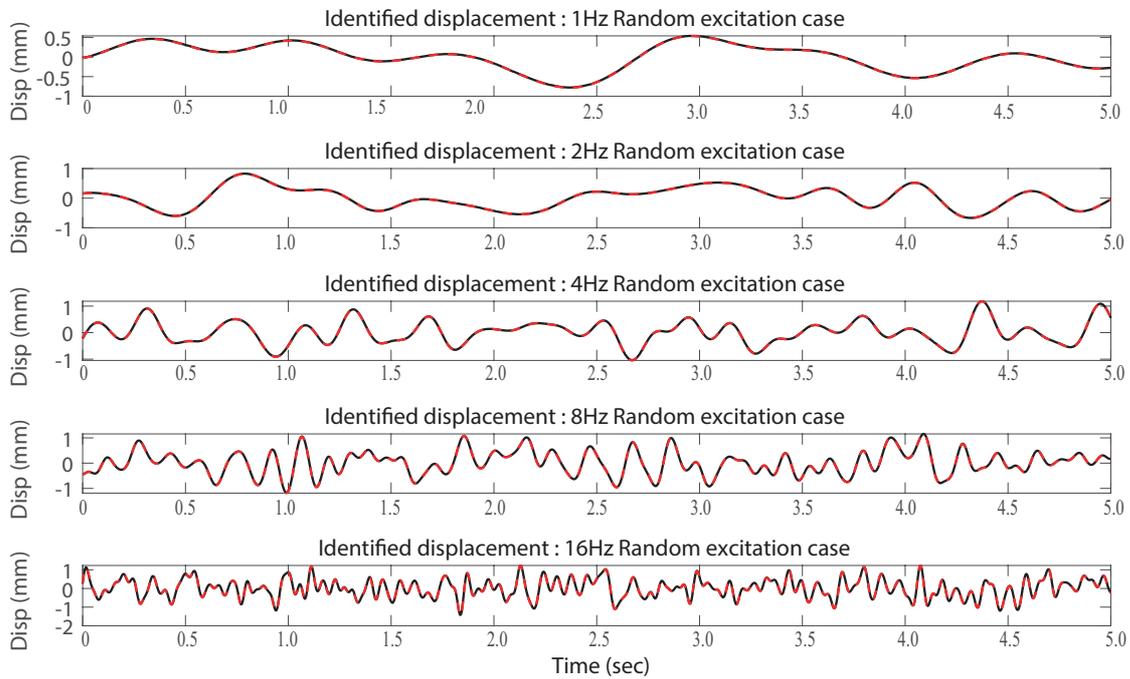}
    \captionof{figure}{Identified displacement response under random loading condition.}    
    \label{fig:Displacement_Random}
    \end{figure}

\begin{figure}[H]
    \centering
    \includegraphics[width=0.5\textwidth]{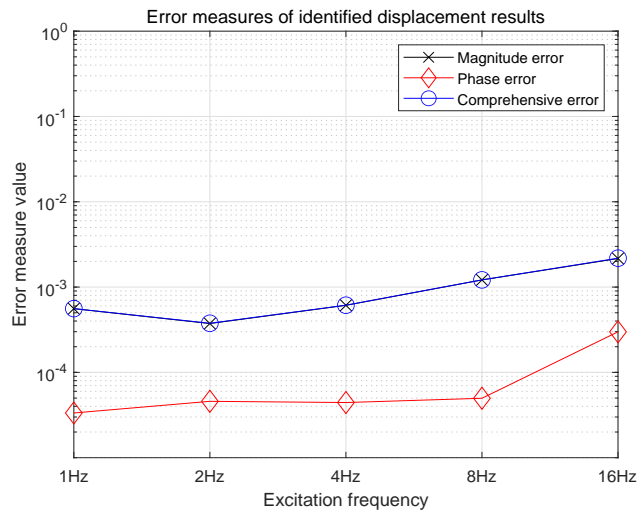}
    \captionof{figure}{Computed error measures for each identified force result}
\label{fig:error_random_disp_exp}
\end{figure}

The identified displacement signals are also evaluated under random excitation conditions.  Figures.~\ref{fig:Displacement_Random} and ~\ref{fig:error_random_disp_exp} show their identified results and computed error measures.
    
    The identified displacement results accurately represent the directly measured displacement signals under random excitation environments. The computed error measures indicate small values under 2.772E-02.

   The required computation times of the test cases are recorded and compared with the actual duration of the experiment. Table.~\ref{table:duration} shows the required computation time results for each case and the average of the values.

%
%
\begin{table}[H]
    \centering
    \caption{Required computation time for identification}
    \begin{tabular}{c|cc}
    \hline
     & \multicolumn{2}{c}{Required time for computation ($\si{\sec}$)}\\
     \hline
    Bandwidth & Sinusoidal exciation & Random excitation\\
    \hline
    1 $\si{Hz}$ & 5.7791 & 5.6816 \\
    2 $\si{Hz}$ & 5.8343 & 5.8862 \\ 
    4 $\si{Hz}$ & 5.8113 & 5.8015 \\ 
    8 $\si{Hz}$ & 5.7859 & 5.7898 \\
    16 $\si{Hz}$ & 5.6424 & 5.9859 \\
    \hline
    Average required time & 5.7706  & 5.829\\
    \hline
    Real duration & \multicolumn{2}{c}{30}\\
    \hline
    \end{tabular}
    \label{table:duration}
\end{table}
    
    As shown in Table.~\ref{table:duration}, the required computation times for every case are much less than the actual duration of the experiment.
    
%
%
\section{Conclusions}\label{section5}
In this study, an implicit inverse force identification algorithm based on the Newmark-$\beta$ time integrator is presented for the structural vibration of an elastic structure containing an acoustic liquid fluid. The accuracy and efficiency of the proposed method are evaluated on both numerical and experimental test beds, and the results show good agreement with reference force information. The displacement response of the structure and the identified forces are also well predicted in the proposed process. The required identification duration of the proposed method is measured while performing the evaluation test and compared with the actual duration of the experiment, which implies that in situ real-time force identification and response reconstruction of the vibroacoustic structure are possible with the proposed method. In particular, from a computational perspective, the proposed method has comparative robustness and computational efficiency compared to the AKF approach in the vibroacoustic problems considered in this work.

Only the vibroacoustic ($u$-$p$) formulation of certain tanker and/or pipe structures with fully filled liquid fluid is considered here, but the proposed inverse force identification method is not limited. The proposed force identification method can also be employed for a tanker with a free surface that has a similar ($u$-$p$) formulation~\cite{KHLEE2022,Ohayonbook}, and it can be derived from different vibroacoustic formulations, such as displacement ($u$) - potential ($\phi$) and displacement ($u$) - pressure ($p$) - potential ($\phi$) forms~\cite{Ohayonbook,kim2020multiphysics}.  
The strongly coupled multiphysics model reduction method \cite{kim2019strongly,kim2020multiphysics} is employed here, but various projection-based model reduction techniques (such as proper orthogonal decomposition modes, coupled modes, and Krylov bases) can be directly employed in the proposed formulation. Finally, this work could be used for the development of real-time monitoring systems and model-driven digital twins for vibroacoustic engineering problems. However, the fluid domain and fluid-induced vibration have received less attention here, and this work does not cover nonlinear responses such as geometrical nonlinearity, the sloshing effect, free surface tension, and prestressed structures. These challenges will be investigated in the future.

%
%
\section*{ Acknowledgements }
This research was funded by National Research Foundation of Korea (NRF-2020M2C9A1062790 and NRF-2021R1A2C4087079).


%
%
\bibliography{references}

\begin{appendix}
\section{Strongly coupled vibro-acroustic model order reduction} \label{appendix:A}

To derive the strongly coupled vibroacoustic reduction formulation, the static relationship between displacement and pressure can be derived by recalling Eq.~\eqref{eq:eom} and assuming that no external forces act on the structural part of the system as follows:

\begin{equation}\label{eq:coupling_relationship}
    \mathbf{u}=\boldsymbol{\Psi}\mathbf{p},~\boldsymbol{\Psi}=[\mathbf{K}^{s}]^{-1}\mathbf{C}.
\end{equation}

To consider the dynamic properties of the structural domain, the displacement vector of the system can be approximated by adding the dynamic mode effect of the structure field, as follows:

\begin{equation}\label{eq:approx_str}
    \hat{\mathbf{u}}\approx\boldsymbol{\Phi}_{d}\mathbf{q}_{d}+\boldsymbol{\Psi}\mathbf{p},
\end{equation}

where $\mathbf{q}$ and $\mathbf{\Phi}$ are the generalized modal response vector of the structure field and its corresponding eigenvector matrix, respectively. Subscript d denotes the chosen dominant modes for the reduction process. 

The mode shape matrix $\mathbf{\Phi}$ can be obtained by solving the eigenvalue problem of the structural domain as follows:

\begin{equation}\label{eq:modal_ind}
        \mathbf{K}^{s}\boldsymbol{\phi}_{i}=\lambda_{i}\mathbf{M}^{s}\boldsymbol{\phi}_{i}~,~i=1,2,\dotsm{N}^{s},
\end{equation}
where $\lambda$ is the eigenvalue of the corresponding eigenmode. Similar to the structural domain, the pressure vector $\mathbf{p}$ can be approximated by decomposing the vector using the eigenvector of the fluid domain. Furthermore, the coupling effects between the structural and fluid domains can be more effectively considered by applying the following approximation relationship:

\begin{equation}\label{eq:approx_str}
    \hat{\mathbf{u}}\approx\boldsymbol{\Phi}_{d}\mathbf{q}_{d}+\boldsymbol{\Psi}\tilde{\boldsymbol{\Xi}}_{d}\tilde{\mathbf{r}}_{d},
\end{equation}

where $\tilde{\mathbf{r}}$ and $\tilde{\mathbf{\Xi}}$ are the generalized response vector of the fluid domain and its eigenvector matrix. Eigenvector can be obtained by solving the following eigenvalue problem:

\begin{equation}\label{eq:evp_fluid}
    \mathbf{K}^{f}\tilde{\xi}_{j}=\tilde{\gamma}_{j} \tilde{\mathbf{M}}^{f} \tilde{\xi}_{j}~,~j=1,2,\dotsm{N}^{f},
\end{equation}

where $\tilde{\xi}$ and $\tilde{\gamma}$ obtained eigenvector and eigenvalue of the system, respectively. $\mathbf{K}^{f}$ is the stiffness matrix of the system and $\tilde{\mathbf{M}}$ is the mass matrix of the partially reduced system, which can consider the strong coupling effects between the structural and fluid domains. From the strongly coupled vibroacoustic reduction technique, the fluid part of the partially reduced mass matrix can be derived as

\begin{equation}\label{eq:eom_partial_red}
    \tilde{\mathbf{M}}^{f}=\mathbf{M}^{f}+\left [\rho_{f} c^{2}\mathbf{C}^{T}+\mathbf{\Psi}^{T}\mathbf{M}^{s}\right]\boldsymbol{\Psi}.
\end{equation}

Hence, the final form of the strongly coupled vibroacoustic reduction technique can be defined as

\begin{equation}\label{eq:Trans_total}
    \hat{\mathbf{T}} \left [ \begin{array}{cc}
        \boldsymbol{\Phi}_{d} & \boldsymbol{\Psi}\tilde{\boldsymbol{\Xi}}_{d} \\
        \mathbf{0} & \tilde{\boldsymbol{\Xi}}_{d}
    \end{array} \right].
\end{equation}

\section{Implementation of the augmented Kalman filter method on the vibroacoustic system}\label{appendix:B}
The method proposed in this work can efficiently identify the unmeasured applied forces and their responses to the vibroacoustic system. A comparative study using the augmented Kalman filter method\cite{lourens2012augmented} is conducted to evaluate the efficiency and accuracy of the proposed method. However, the original work of the augmented Kalman filter only considered structural dynamic systems, and the reduced-order modeling method was not implemented. Hence, to perform a comparative study under equivalent conditions, the formulation of the original augmented Kalman filter is redefined based on the reduced-order model of the vibroacoustic system. From the reduced form of the vibroacoustic system, the equations of motion can be described as
\begin{equation}
    \hat{\mathbf{A}}\ddot{\hat{\mathbf{d}}}+\hat{\mathbf{D}}\dot{\hat{\mathbf{d}}}+\hat{\mathbf{B}}{\hat{\mathbf{d}}}=\hat{\mathbf{T}}^{T}\mathbf{S}_{f}\mathbf{f}
\end{equation}

The equations of motion can be reorganized in a state-space form as

\begin{subequations}
\begin{align}
    \left[ \begin{array}{c} \dot{\hat{\mathbf{d}}} \\ \ddot{\hat{\mathbf{d}}} \end{array} \right]&=\mathbf{A}_{c}\left[ \begin{array}{c} \hat{\mathbf{d}} \\ \dot{\hat{\mathbf{d}}} \end{array} \right] + \mathbf{B}_{c}\mathbf{f}\\
    \mathbf{A}_{c}&=\left[ \begin{array}{cc}
        \mathbf{0} & \mathbf{I}  \\
        -\hat{\mathbf{A}}^{-1}\hat{\mathbf{B}} & -\hat{\mathbf{A}}^{-1}\hat{\mathbf{D}}
    \end{array} \right],~
    \mathbf{B}_{c}=\left[ \begin{array}{c}
         \mathbf{0}\\
         \hat{\mathbf{A}}^{-1}\hat{\mathbf{T}}^{T}\mathbf{S}_{f}
    \end{array} \right]
    \end{align}
\end{subequations}

The defined state-space equation can then be discretized in the time domain using the exponential time integration method, as follows:
 
\begin{equation}
    \left[ \begin{array}{c} {}^{t+\Delta t}\hat{\mathbf{d}} \\ {}^{t+\Delta t}\dot{\hat{\mathbf{d}}} \end{array} \right]= e^{\mathbf{A}_{c}\Delta t} \left[ \begin{array}{c} {}^{t}\hat{\mathbf{d}} \\ {}^{t}\dot{\hat{\mathbf{d}}} \end{array} \right]+\left[ \left[\mathbf{A}_{c}- \mathbf{I} \right] \mathbf{A}_{c}^{-1}\mathbf{B}_{c} \right] {}^{t}\mathbf{f}
\end{equation}

To identify the applied forces and their responses concurrently, the augmented form of the state-space equation can be redefined as
 
\begin{subequations}
\begin{align}
    {}^{t+\Delta t}\mathbf{x}&=\mathbf{A}_{a}{}^{t}\mathbf{x}\\
    \mathbf{A}_{a}&=\left[ \begin{array}{cc} e^{\mathbf{A}_{c}\Delta t} & \left[ \mathbf{A}_{c}-\mathbf{I} \right] \mathbf{A}_{c}^{-1}\mathbf{B}_{c} \\ \mathbf{0} & \mathbf{I} \end{array} \right]~,~
    \mathbf{x}=\left[\begin{array}{c} \hat{\mathbf{d}} \\ \dot{\hat{\mathbf{d}}} \\\mathbf{f} \end{array}\right]
\end{align}
\end{subequations}

Meanwhile, the relationship between the measured response signal and numerical prediction can be defined as

\begin{subequations}
\begin{align}
    {}^{t}\mathbf{z}&=\mathbf{G}_{a}{}^{t}\mathbf{x}\\
    \mathbf{G}_{a}&=\left[ \begin{array}{ccc} \mathbf{S}_{d}\hat{\mathbf{T}}-\mathbf{S}_{a}\hat{\mathbf{T}}\hat{\mathbf{A}}^{-1}\mathbf{B} & \mathbf{S}_{v}\hat{\mathbf{T}}-\mathbf{S}_{a}\hat{\mathbf{T}}\hat{\mathbf{A}}^{-1}\hat{\mathbf{D}} & \mathbf{S}_{a}\hat{\mathbf{T}}\hat{\mathbf{A}}^{-1}\hat{\mathbf{T}}^{T}\mathbf{S}_{f} \end{array} \right].
\end{align}
\end{subequations}

The transformation matrix $\hat{\mathbf{T}}$ is applied to define the relationship between the measured physical quantities and the reduced response vector. By including the process and measurement noises ${}^{t}\boldsymbol{\zeta}$ and ${}^{t}\mathbf{v}$, the augmented state-space model of the method can then be formulated as

\begin{subequations}
\begin{align}
    {}^{t+\Delta t}\mathbf{x}=\mathbf{A}_{a}{}^{t}\mathbf{x}+{}^{t}\boldsymbol{\zeta}\\
    {}^{t}\mathbf{z}=\mathbf{G}_{a}{}^{t}\mathbf{x}+{}^{t}\mathbf{v}
\end{align}
\end{subequations}

By implementing the augmented Kalman filter algorithm on the derived model, the following inverse force identification algorithm can be derived:

Measurement update:
\begin{subequations}
\begin{align}
{}^{t}\mathbf{L}&={}^{t|t-\Delta t}\mathbf{P}\mathbf{G}_{a}^{T}\left( \mathbf{G}_{a}{}^{t|t-\Delta t}\mathbf{P}\mathbf{G}_{a}^{T}+\mathbf{R} \right)^{-1},\\
{}^{t|t}\mathbf{x}&={}^{t|t-\Delta t}\mathbf{x}+{}^{t}\mathbf{L}\left( {}^{t}\mathbf{z}-\mathbf{G}_{a}{}^{t|t-\Delta t}\mathbf{x} \right),\\
{}^{t|t}\mathbf{P}&={}^{t|t-\Delta t}\mathbf{P}-{}^{t}\mathbf{L}\mathbf{G}_{a}{}^{t|t-\Delta t}\mathbf{P},
\end{align}
\end{subequations}

Time update:
\begin{subequations}
\begin{align}
    {}^{t+\Delta t|t}\mathbf{x}&=\mathbf{A}_{a}{}^{t|t}\mathbf{x},\\
    {}^{t+\Delta t|t}\mathbf{P}&=\mathbf{A}_{a}{}^{t|t}\mathbf{P}\mathbf{A}_{a}^{T}+\mathbf{Q},
\end{align}
\end{subequations}
\end{appendix}
where $\mathbf{Q}$ and $\mathbf{R}$ are the covariance matrices of the process and measurement errors, respectively; and $\mathbf{P}$ is the identification covariance error matrix. The applied forces and their responses to the vibroacoustic systems can be identified by iteratively implementing the redefined process.

\end{document}